\documentclass[10pt,journal,spssoc]{IEEEtran}

\ifCLASSOPTIONcompsoc
  \usepackage[nocompress]{cite}
\else
  \usepackage{cite}
\fi

\ifCLASSINFOpdf
\else
\fi

\usepackage{diagbox}
\usepackage{amssymb}
\usepackage{amsmath}
\usepackage{mathtools}
\usepackage[templates]{genealogytree}
\usepackage[utf8]{inputenc}
\usepackage{adjustbox}
\usepackage[T1]{fontenc}
\usepackage{todonotes}
\usepackage{array}
\usepackage[english]{babel}
\usepackage{blindtext}
\usepackage{url}
\usepackage{listings}
\usepackage{makecell}
\usepackage{multirow}
\usepackage[makeroom]{cancel}
\usepackage[normalem]{ulem}
\usepackage{placeins}
\usetikzlibrary{shapes}
\usepackage{color}

\newcolumntype{x}[1]{>{\centering\arraybackslash}p{#1}}
\newcolumntype{N}{>{\centering\arraybackslash}m{.5in}}
\newcolumntype{G}{>{\centering\arraybackslash}m{2in}}
\newcolumntype{p}[1]{>{\centering\arraybackslash}p{#1}}

\usepackage{ifthen}
\newboolean{showcomments}
\setboolean{showcomments}{true} 
\ifthenelse{\boolean{showcomments}}{
\newcommand{\mynote}[3]{\noindent{\color{#3}\textbf{#1:\xspace} #2}}}
{ \newcommand{\mynote}[3]{}
}
\definecolor{mypurple}{rgb}{0.6,0.4,0.8}
\newcommand{\hh}[1]{\mynote{hh}{#1}{blue}}
\newcommand{\as}[1]{\mynote{as}{#1}{blue}}

\renewcommand{\hh}[1]{}
\usepackage{hyperref}

\usepackage{ulem,lipsum}
\newcommand\soutpars[1]{\let\helpcmd\sout\parhelp#1\par\relax\relax}
\long\def\parhelp#1\par#2\relax{
  \helpcmd{#1}\ifx\relax#2\else\par\parhelp#2\relax\fi
}
\usetikzlibrary{calc}
\usepackage[noend]{algorithmic}
\usepackage[algoruled,nofillcomment,algo2e]{algorithm2e}
\newenvironment{protocol}[1][htb]
  {
   \begin{algorithm2e}[#1]
  }{\end{algorithm2e}}

\usepackage{booktabs}
\usepackage{tikz,pgfplots}
\pgfplotsset{ytick style={draw=none}}
\pgfplotsset{xtick style={draw=none}}
\usepackage{siunitx}
\pgfplotsset{width=7cm,compat=1.3}
\usepackage{pgfplotstable}

 \usetikzlibrary{positioning,shadows}
\usepackage[caption=false,font=footnotesize]{subfig}
\tikzstyle{chart}=[
legend label/.style={font={\scriptsize},anchor=west,align=left},
legend box/.style={rectangle, draw, minimum size=5pt},
axis/.style={black,semithick,->},
axis label/.style={anchor=east,font={\tiny}},
]
\tikzstyle{bar chart}=[
chart,
bar width/.code={
	\pgfmathparse{##1/2}
	\global\let\bar@w\pgfmathresult
},
bar/.style={very thick, draw=white},
bar label/.style={font={\bf\small},anchor=north},
bar value/.style={font={\footnotesize}},
bar width=.75,
]
\DeclareMathOperator*{\argmin}{arg\,min}
\definecolor{darkgreen}{RGB}{47,109,79}
\definecolor{darkblue}{RGB}{57,79,99}
\definecolor{rosso}{RGB}{220,57,18}
\definecolor{giallo}{RGB}{255,153,0}
\definecolor{blu}{RGB}{102,140,217}
\definecolor{verde}{RGB}{16,150,24}
\definecolor{viola}{RGB}{153,0,153}
\definecolor{awesome}{rgb}{1.0, 0.13, 0.32}
\definecolor{ref}{rgb}{0.65,0.65,0.65} 
\definecolor{darkgreen}{RGB}{47,109,79}
\definecolor{darkblue}{RGB}{57,79,99}
\definecolor{rosso}{RGB}{220,57,18}
\definecolor{verde}{RGB}{16,150,24}
\definecolor{viola}{RGB}{153,0,153}
\definecolor{americanrose}{rgb}{1.0, 0.01, 0.24}
\definecolor{bostonuniversityred}{rgb}{0.8, 0.0, 0.0}
\definecolor{chocolate(traditional)}{rgb}{0.48, 0.25, 0.0}
\definecolor{violet(web)}{rgb}{0.93, 0.51, 0.93}
\definecolor{airforceblue}{rgb}{0.36, 0.54, 0.66}

\definecolor{almond}{rgb}{0.94, 0.87, 0.8}
\definecolor{amethyst}{rgb}{0.6, 0.4, 0.8}
\definecolor{bazaar}{rgb}{0.6, 0.47, 0.48}
\definecolor{britishracinggreen}{rgb}{0.0, 0.26, 0.15}
\definecolor{byzantine}{rgb}{0.74, 0.2, 0.64}
\definecolor{cadetblue}{rgb}{0.37, 0.62, 0.63}
\definecolor{cambridgeblue}{rgb}{0.64, 0.76, 0.68}
\definecolor{candypink}{rgb}{0.89, 0.44, 0.48}
\definecolor{caputmortuum}{rgb}{0.35, 0.15, 0.13}
\definecolor{cerulean}{rgb}{0.0, 0.48, 0.65}
\definecolor{corn}{rgb}{0.98, 0.93, 0.36}
\definecolor{darkbyzantium}{rgb}{0.36, 0.22, 0.33}
\definecolor{darkgoldenrod}{rgb}{0.72, 0.53, 0.04}
\definecolor{darkseagreen}{rgb}{0.56, 0.74, 0.56}
\definecolor{darkturquoise}{rgb}{0.0, 0.81, 0.82}
\definecolor{dimgray}{rgb}{0.41, 0.41, 0.41}
\definecolor{eggplant}{rgb}{0.38, 0.25, 0.32}

\usetikzlibrary{patterns}

\usepackage{amssymb}
\usepackage{amsmath}
\usepackage{array}

\begin{document}
\title{PrivEdge: From Local to Distributed Private Training and Prediction}

\author{Ali Shahin Shamsabadi, Adri\`a Gasc\'on, Hamed Haddadi and Andrea Cavallaro
\IEEEcompsocitemizethanks{\IEEEcompsocthanksitem Ali  Shahin  Shamsabadi  and  Andrea  Cavallaro  are  with Centre for Intelligent Sensing, Queen Mary University of London. Adri\`a Gasc\'on is with The Alan Turing Institute. Hamed Haddadi is with the Faculty of Engineering, Imperial College London. This work was partly done during an Enrichment scheme at The Alan Turing Institute. Adri\`a Gasc\'on was supported by The Alan Turing Institute under the EPSRC grant EP/N510129/1, and funding from the UK Government’s Defence \& Security Programme in support of the Alan Turing Institute. Hamed Haddadi was partially supported by the EPSRC Databox and DADA grants (EP/N028260/1, EP/R03351X/1) and a gift from Huawei Technologies. Andrea Cavallaro wishes to thank The Alan Turing Institute (EP/N510129/1), which is funded by the EPSRC, for its support through the project PRIMULA. Copyright 2020 IEEE.  Personal use of this material is permitted.  Permission from IEEE must be obtained for all other uses, in any current or future media, including reprinting/republishing this material for advertising or promotional purposes, creating new collective works, for resale or redistribution to servers or lists, or reuse of any copyrighted component of this work in other works. \protect\\
}
}

\IEEEtitleabstractindextext{
\begin{abstract}
Machine Learning as a Service (MLaaS) operators provide model training and prediction on the cloud. MLaaS applications often rely on centralised collection and aggregation of user data, which could lead to significant privacy concerns when dealing with sensitive personal data. To address this problem, we propose PrivEdge, a technique for privacy-preserving MLaaS that safeguards the privacy of users who provide their data for training, as well as users who use the prediction service. With PrivEdge, each user independently uses their private data to locally train a \emph{one-class reconstructive adversarial network} that succinctly represents their training data. As sending the model parameters to the service provider in the clear would reveal private information, PrivEdge secret-shares the parameters among two non-colluding MLaaS providers, to then provide cryptographically private prediction services through secure multi-party computation techniques. We quantify the benefits of PrivEdge and compare its performance with state-of-the-art centralised architectures on three privacy-sensitive image-based tasks: individual identification, writer identification, and handwritten letter recognition. Experimental results show that PrivEdge has high precision and recall in preserving privacy, as well as in distinguishing between private and non-private images. Moreover, we show the robustness of PrivEdge to image compression and biased training data. The source code is available at {\url{https://github.com/smartcameras/PrivEdge}}.

\end{abstract}

}

\maketitle

\IEEEdisplaynontitleabstractindextext

\IEEEpeerreviewmaketitle

\ifCLASSOPTIONcompsoc
\IEEEraisesectionheading{\section{Introduction}\label{sec:introduction}}
\else
\section{Introduction}
\label{sec:introduction}
\fi

\IEEEPARstart{M}{achine} Learning as a Service (MLaaS) is increasingly adopted in a range of applications including authentication through signature~\cite{sae2014online}, access control through face recognition~\cite{MicrosoftAzure, MicrosoftAzureDoor}, and annotation of pictures in social media~\cite{choi2011collaborative}. The need to train machine learning models from user data, such as images, increases the risk of privacy violations~\cite{facebook,qiu2016survey}. A privacy violation may arise when a service provider or a user learns private information of another user from the data or model parameters during either training or prediction.  

The training of MLaaS applications can be centralised~\cite{sweeney2002k,malekzadeh2018protecting} or distributed~\cite{shokri2015privacy,mcmahan2016communication,mcmahan2017learning,Shahin2018, aono2017privacy}. In {\em centralised learning}, the service provider collects all users' data to train a model. To preserve privacy, local data anonymisation~\cite{sweeney2002k,malekzadeh2018protecting} or data encryption~\cite{hesamifard2018privacy} can be performed on the user side, prior to transmission to the service provider. However, the curse of dimensionality may render these approaches impractical~\cite{osia2017hybrid}. In addition to this, encryption may be needed to approximate the model, thus reducing accuracy~\cite{hesamifard2018privacy}. In {\em distributed learning}, the service provider trains a model by iteratively aggregating the parameters or gradients of models trained locally by users~\cite{shokri2015privacy,mcmahan2016communication,mcmahan2017learning}. If the models trained on private data are shared among users or with the service provider in the clear, the privacy of each user in the training may be violated by other users who may be honest-but-curious (i.e.~honest in performing the operations, but also attempting to learn information about other users by analysing the received information~\cite{aggarwal2008general}) or by an {honest-but-curious} service provider during training or prediction~\cite{aono2017privacy,hitaj2017deep,zhang2016understanding,melis2018inference}. Finally, in distributed learning methods~\cite{shokri2015privacy,mcmahan2016communication,mcmahan2017learning, aono2017privacy}, users are assumed to have access to the data of several classes with machine learning models that forget previously learned information upon learning new information (catastrophic forgetting)~\cite{french1999catastrophic,mccloskey1989catastrophic}. 

Distributed one-class learning addresses {honest-but-curious} users and catastrophic forgetting issues in applications such as image-sharing social media~\cite{facebook}, where each user has data of one class (e.g. their face), by decomposing the multi-class classifier into a set of one-class classifiers, which are trained locally~\cite{Shahin2018}. However, users need to train their classifier using a public feature extractor, trained by the service provider, which could use it to extract private information. Moreover, the service provider needs to collect data for training the feature extractor, which may be difficult when dealing with private data. More importantly, the service provider has access to the parameters of all local models as they are uploaded in the clear.

To address these problems and enable privacy-preserving MLaaS, we propose \emph{PrivEdge}, a technique that not only preserves the privacy of users participating in the training, but also the privacy of users that submit their data to a prediction service. The users and the service provider do not exchange data nor model parameters in the clear during training or prediction\footnote{Note that PrivEdge reveals the classification output and therefore does not protect against  membership inference attacks~\cite{shokri2017membership} and model extraction~\cite{tramer2016stealing}.}. We model each user as a distinct class and propose one-class \emph{Reconstructive Adversarial Networks} (RANs) that enable users to perform {\em private training} locally without a public pre-trained feature extractor, unlike recent distributed learning work~\cite{Shahin2018}.

A RAN reconstructs private images via an autoencoder aided by a discriminator network trained to differentiate between output and input of the autoencoder.  Importantly, a user can join the framework at any time by training their one-class RAN.  
For {\em private prediction}, we assume that a non-colluding regulator is available and use the $2$-server model of Multi-Party Computation (MPC)~\cite{mohassel2017secureml}. A 2-Party Computation (2PC) protocol\footnote{Yao's millionaires' problem~\cite{yao1982protocols} is an example of secure 2PC wanting to determine which party is richer, without disclosing to each other their wealth (i.e.~values of their input).} is secure if the two participating parties can {\em only} learn something that can be computed based on their input and output. Then, the regulator aids in the computation while learning nothing from the private data~\cite{mohassel2017secureml,gascon2017privacy,nikolaenko2013privacy,nikolaenko2013privacym}. In our scenario (inference of one-class classifiers), two parties either compute a multiplication between layers, or do a comparison for computing the activation function\footnote{The same as the one in the millionaires' problem~\cite{yao1982protocols}.}. Our 2PC protocol involves a Yao's garbled circuit implementation of the Leaky Rectifier Linear Unit (L-ReLU) activation function~\cite{glorot2011deep,maas2013rectifier}. 

The remainder of the paper is organised as follows. We begin by describing methods that can provide private training or prediction in Section~\ref{sec:background}. In Section~\ref{sec:problem}, we describe the problem followed by discussing our setting and threat model. In Sections~\ref{sec:train} and~\ref{sec:prediction}, we describe in detail the private training and prediction of PrivEdge. In Section~\ref{sec:result}, we discuss the evaluation on three real-world privacy-sensitive datasets. Finally, we conclude in Section~\ref{sec:conclude}.

\section{Background}
\label{sec:background}

Private training and prediction for MLaaS applications can be addressed by applying cryptography tools such as Fully Homomorphic Encryption (FHE)~\cite{gentry2009fully} and Multi-Party Computation (MPC)~\cite{goldreich1998secure}, as well as information theory and differential privacy~\cite{dwork2006our}.

FHE~\cite{gentry2009fully} encrypts data to enable anyone to perform arbitrary functions on the encrypted data without compromising the encryption. FHE is computationally expensive and, despite recent advances~\cite{gentry2012fully,gentry2013homomorphic,ducas2015fhew,chillotti2016faster}, training a machine learning classifier (e.g.~a Deep Neural Network, DNN) using FHE is still unfeasible. A weaker (i.e.~partially homomorphic) version of FHE, Additively Homomorphic Encryption (AHE), was used for distributed learning to protect the parameters via encryption before sending them to the service provider~\cite{aono2017privacy}. When the encryption scheme is {\em homomorphic} to addition, the service provider can add, without decryption, encrypted parameters of all users and fine-tune the global DNN classifier~\cite{aono2017privacy}. Users can then decrypt the global encrypted parameters to fine-tune them locally on their data. Trusting the users is a critical assumption in this approach, as intermediate results in the computation (i.e.~the  gradients) are revealed to the users. Moreover, an honest-but-curious service provider could span a fake user to achieve the same goal. Finally, the parameters for the encryption must be generated by a party that does not collude with the service provider, to prevent the service provider accessing the parameters in the clear. CryptoNets~\cite{gilad2016} and TAPAS~\cite{sanyal2018tapas} do private prediction using homomorphic encryption. Apart from their expensive computations, CryptoNets and TAPAS may negatively affect the accuracy of the prediction by changing the structure of the model, replacing the non-linear activation functions (e.g.~Rectifier linear unit, and Sigmoid) with polynomial functions in the CryptoNets, and binarising the parameters in TAPAS. 

The ability of a service provider to access private data beyond what the model outputs can be limited using alternative approaches based on MPC~\cite{goldreich1998secure}. MPC has scalability issues, as it needs a large number of communication rounds to run interactive protocols between each pair of parties for each gate in the circuit\footnote{Consider representing functions with arithmetic or Boolean circuit including gates.}~\cite{hazay2010efficient}. To address these issues, the so-called 2-server MPC model can be used~\cite{mohassel2017secureml,gascon2017privacy,nikolaenko2013privacy,nikolaenko2013privacym}. The 2-server MPC model involves a third-party (e.g. another service provider) that is trusted not to collude with the service provider when performing the training procedure in a cryptographically secure 2PC protocol. In practice, collusion does not occur, especially in cloud computing, as service providers do not have incentive for cooperation~\cite{kamara2011outsourcing}. Collusion is either too costly (not feasible) for service providers, or prevented by regulations or physical means (e.g.~ballot boxes~\cite{lepinksi2005collusion}, mediator model~\cite{alwen2008collusion}). The training is run as a secure protocol by modifying DNNs and approximating their non-linear layers with polynomial functions, thus negatively impacting accuracy. Although MPC is faster than FHE, training is still time consuming. For example, the offline phase of SecureML~\cite{mohassel2017secureml} for a simple fully-connected DNN with $2$ hidden layers takes $\sim$3 days on two Amazon EC2 c4.8xlarge machines running Linux. The prediction, however, is significantly faster, requiring less than $5$ seconds in total. For these reasons, MPC techniques are used for private prediction~\cite{malik2016uses, liu2017oblivious,juvekar2018gazelle,rouhani2017}, but the cost of communication between parties remains as a major bottleneck.

In contrast to FHE and MPC, which fully encrypt the data to request a prediction, users could remove the private information from data and only send extracted features to the service provider. A feature extractor can be obtained by solving an optimisation problem that maximises the mutual information between the feature set and non-sensitive variables, while minimising mutual information between the feature set and private variables~\cite{osia2018deep,osia2017hybrid,osia2018private}.

Differential privacy~\cite{dwork2006our, dwork2008differential,wasserman2010statistical,dwork2006calibrating,mcsherry2007mechanism} can obfuscate information that determine the presence (or absence) of data about an individual in the training dataset during training or prediction. The prediction of differentially private distributed learning~\cite{mcmahan2017learning} is private as users can query the differentially private model locally. However, applying differential privacy negatively affects accuracy~\cite{bagdasaryan2019differential}.

In summary, state-of-the-art private training methods using FHE, MPC or differential privacy have high computational and communications costs, and reduce the model accuracy~\cite{bagdasaryan2019differential}. We tackle these issues by ensuring that the training can be done locally by each user in the clear using our proposed one-class classifier, which does not require interaction with other users. We use MPC techniques, which have significant overhead, only for prediction and to avoid revealing parameters of the trained model and the users’ query data to the service provider.

\section{Problem description and Threat model}
\label{sec:problem}
In this section we formulate the problem, describe our setting and the capabilities of the adversaries that PrivEdge can withstand.

Let us define an image as {\em private} if it includes content such as the face or handwritten text of an individual. Let us consider the MLaaS task of predicting whether an image $\mathbf{X}$ submitted to the service provider by each user $u_n$ is an image with private content of user $u_i$, where $i=1, ..., N$, $i \neq n$, and $N$ is the total number of users who decided to protect their private images.  

The service provider aims to build an $N$-class classifier using training data contributed by $N$ users, to decide whether a new image belongs to one of the private classes. Let 
\begin{equation*}
  X_i=\{\mathbf{X}_{i,1},...,\mathbf{X}_{i,j},...,\mathbf{X}_{i,K_i}\},  
\end{equation*}
be a set of private images of user $u_i$,  where $ K_i \in \mathbb{N}$, and
\begin{equation*}
    \hat{X}_i=\{\mathbf{\hat{X}}_{i,1},...,\mathbf{\hat{X}}_{i,j},...,\mathbf{\hat{X}}_{i,\hat{K}_i}\},
\end{equation*}
be a set of non-private images, where $ \hat{K}_i \in \mathbb{N}$. The cardinality of the two sets, $K_i$ and $\hat{K}_i$, may differ considerably across users.

To avoid disclosing private images of a user or the model parameters of the classifier to other users or to the service provider, the training phase of the multi-class classifier must be private. The privacy of the users must also be preserved in the prediction phase, by keeping the model and images private until they are classified as non-private and, for example, shared on  social media. In the case of privacy-preserving image-sharing MLaaS, users shall be able to control what images about themselves are shared by others. In this context, preserving the privacy of a user means preventing other users from uploading private images of others. In this case, classes are disjoint and private across users.  Moreover, the ideal solution should prevent the service provider and other users from learning private information from images of a user\footnote{Consider for example the case of Facebook that asked users for their private images to use as training set~\cite{facebook}.}.  

In PrivEdge, $N$ users $u_1,...,u_n,...,u_N$ train the model, while the service provider and the regulator do the prediction without further involvement of the users in the computations. 

In our threat model, an arbitrarily large number of parties including the service provider and users might collude to try to recover information about a particular user during training. Note that we do not put any computational restrictions on such an adversary. This threat model is too strong to be satisfied using MPC techniques, as it requires information theoretic guarantees under unbounded collusions~\cite{benaloh1986secret}.

We address this strong threat model by ensuring that the training phase is performed locally by each user, and such local computation does not involve data from other users. Therefore, most of the complexity of PrivEdge is in the prediction phase, as in this phase the data from different users is combined.

In the prediction phase, we rely on an external honest-but-curious party, which we call regulator. The regulator does not collude with the service provider and aids in the computation without accessing input data. In the prediction phase, we assume honest-but-curious parties and arbitrary collusions, except for the regulator and the service provider (as stated in Section~\ref{sec:introduction}). This enables the so-called 2-server model of MPC~\cite{mohassel2017secureml,gascon2017privacy,nikolaenko2013privacy,nikolaenko2013privacym}. As above, we model privacy in a secure computation sense, by enforcing that nothing but the result of the computation is revealed. This is formalised in the simulation framework of MPC~\cite{goldreich1998secure}. Hence, in our setting the computation must reveal the predicted class or a special unknown class, and nothing else.

In PrivEdge, we achieve protection against the adversaries defined above by ensuring that an honest-but-curious user, who can collude with any subsets of the users and at most one of two service providers, learns nothing about the inference of the one-class classifiers beyond those classifiers and the assigned classes corresponding to the users with whom they are colluding. 
 
As MPC does not impose any constraints on the parameters provided by users, an honest-but-curious user may learn an identity function (i.e. perfect reconstruction of any input image) with their own one-class classifier in order to assign any image to them. To the best of our knowledge, an identity function can only be learned by an \textit{over-complete} autoencoder in which the dimensions of the hidden layers are equal to or larger than the input and output layers. We design each one-class RAN with an \textit{under-complete} autoencoder, which compresses the images in addition to verifying users' architectures with the service provider. Proving the impossibility of copying the input of the under-complete autoencoder to the output layer is an important avenue for future work.

\section{Private training}
\label{sec:train}

An $N$-class classifier can be decomposed into $N$ classifiers via binary~\cite{galar2011overview} or one-class decomposition~\cite{krawczyk2015usefulness}. A binary decomposition tends to fail when only data of one class are available~\cite{japkowicz2003class, krawczyk2015usefulness, bellinger2012one}. Instead, one-class classifiers (or data descriptors)~\cite{tax2001one} learn  to distinguish the target class from outlier classes (unavailable training subsets) via density estimation~\cite{duda1973pattern}, closed boundary estimation~\cite{tax2004support,tax2003kernel,scholkopf2001estimating} or data reconstruction~\cite{manevitz2007one}. 

We propose to learn the reconstruction descriptor of private images by training a one-class RAN composed of a reconstructor and a discriminator. The one-class RAN reconstructs images of users by minimising reconstruction and adversary losses. While our architecture is inspired by Generative Adversarial Networks (GANs)~\cite{goodfellow2014generative}, 
in the sense that the discriminator is similar to that in GANs, the reconstructor has three major differences from a standard GAN generator. First, the input of the reconstructor is a set of images, as opposed to noise. Second, the reconstructor aims to reconstruct the input, i.e.~making input and output images similar. Third, we replace the negative log-likelihood objectives by a mean square loss for increased stability during training and better quality of the reconstructed images~\cite{mao2016multi}.

We instantiate the discriminator of each user $u_i$, $\mbox{D}_i(\cdot):\mathbb{R}^{w \times h \times c} \rightarrow \{0,1\}^{d}$ (where $w$ is the input width, $h$ is the input height, and $c$ is the number of input channels) with a deep convolutional binary classifier ($d=2$) parametrised by
\begin{equation*}  
    {W}_{\mbox{D}_i} = \{{\mathbf{W}}_{\mbox{D}_i}^{p} \in \mathbb{R}^{w_k \times h_k \times c_p \times c_{p+1}}: p =1,..,P \},
\end{equation*} 
where $w_k$ is the kernel width, $h_k$ is the kernel height, and $c_p$ and $c_{p+1}$ are the number of feature channels in the $p$-th and $(p+1)$-th convolutional layers. Each reconstructor $\mbox{R}_i(\cdot):\mathbb{R}^{w \times h \times c} \rightarrow \mathbb{R}^{w \times h \times c}$ is instantiated with a $T$-layer deep convolutional autoencoder~\cite{vincent2010stacked,shamsabadi2017new} composed of a parametric encoder, followed by a parametric decoder
\begin{equation*}  
{W}_{\mbox{R}_i} = \{{\mathbf{W}}_{\mbox{R}_i}^t \in \mathbb{R}^{w_k \times h_k \times c_t \times c_{t+1}}: t =1,..,T\}.
\end{equation*}

The encoder maps images into a compressed set of features, while the decoder reconstructs the images from the set of features extracted by the encoder.

In the training phase, each user $u_i$ trains their one-class RAN, $\mbox{M}_i=(\mbox{R}_i(\cdot),\mbox{D}_i(\cdot))$, on their private image set, $X_i$, by alternating between learning parameters for the discriminator and the reconstructor until convergence.

To learn the final parameters of $\mbox{R}_i(\cdot)$ in each iteration, ${W}_{\mbox{R}_i}^*$, user $u_i$ fixes the parameters of $\mbox{D}_i(\cdot)$ and minimises the loss function, which includes two terms, reconstruction loss $L_r$ and adversary loss $L_a$, based on ${W}_{\mbox{R}_i}$, using the mini-batch Adam optimiser~\cite{kingma2014adam}:

\begin{equation}
\label{eq:opt-g}
{W}_{\mbox{R}_i}^* = \argmin_{{W}_{\mbox{R}_i}} \Big( \gamma L_r \big(\mathbf{X}_{i,j},\mathbf{\bar{X}}_{i,j}\big) + \beta L_a \big(\mathbf{\bar{X}}_{i,j},\mathbf{y}_f^j\big)\Big), 
\end{equation}
where $\gamma$ and $\beta$ are the hyper-parameters that control the trade-off between the effects of the reconstruction and adversary losses on the parameters of the reconstructor. $\mathbf{\bar{X}}_{i,j}\in \mathbb{R}^{w \times h \times c}$ is the reconstructed image of the same size as the input image $\mathbf{X}_{i,j}$
\begin{equation*}
    \mathbf{\bar{X}}_{i,j} = \mbox{R}_i({W}_{\mbox{R}_i},\mathbf{X}_{i,j}),
\end{equation*}
and $Y_f^j$ denotes their fake labels as:
\begin{equation*}
     Y_f =\{\mathbf{y}_{f}^{j} \in \mathbb{R}^{d} : {j} =1,..,K_i \}.
\end{equation*}

The reconstruction loss, $L_r$, measures how well the private images of each $u_i$ are reconstructed by $\mbox{R}_i(\cdot)$ through computing the mean square error between the input images, $\mathbf{X}_{i,j}$, and the reconstructed images, $\mathbf{\bar{X}}_{i,j}$:

\begin{equation*}
L_r = \frac{1}{K_id} \sum_{j=1}^{K_i} \sum_{l=1}^{d=w \times h \times c} \big(x_{i,j}(l)-\bar{x}_{i,j}(l)\big)^2,
\end{equation*}
where $x_{i,j}(l)$ and $\bar{x}_{i,j}(l)$ are the $l$-th elements of image $\mathbf{{X}}_{i,j}$ and the reconstructed image $\mathbf{\bar{X}}_{i,j}$, respectively. 

The adversary loss, $L_a$, is the mean square error between the label of $\mathbf{\bar{X}}_{i,j}$ predicted by the discriminator, 
\begin{equation*}
\mathbf{\bar{y}}_{p}^{j}=\mbox{D}({W}_{\mbox{D}_i},\mathbf{\bar{X}}_{i,j}),    
\end{equation*}
and the fake label of the reconstructed images, $\mathbf{y}_{f}^{j}$, as:
\begin{equation*}
L_a = \frac{1}{K_i} \sum_{j=1}^{K_i}\Big(\mathbf{\bar{y}}_{p}^{j}-\mathbf{y}_{f}^{j}\Big)^{2}.
\end{equation*}

Hence, assuming a fixed ${W}_{\mbox{D}_i}$ and optimising Equation~\ref{eq:opt-g} enable user $u_i$ to reconstruct the images through $\mbox{R}_i(\cdot)$ in a way that $\mbox{D}_i(\cdot)$ cannot distinguish the reconstructed images from the original images.

The final set of parameters of $\mbox{D}_i(\cdot)$ in this iteration, ${W}_{\mbox{D}_i}^*$, are learned by solving the following optimisation problem based on ${W}_{\mbox{D}_i}$, while ${W}_{\mbox{R}_i}$ are fixed, using a mini-batch Adam optimiser:
\begin{equation}
\label{eq:dis_loss}
{W}_{\mbox{D}_i}^* = \argmin_{{W}_{\mbox{D}_i}} \frac{1}{2K_i}\sum_{j=1}^{K_i}\Big(\big(\mathbf{\bar{y}}_{p}^{j}-\mathbf{y}_{r}^{j}\big)^{2}+ \big(\mathbf{{y}}_{p}^{j}-\mathbf{y}^{j}\big)^{2}\Big),
\end{equation}
where $\mathbf{y}_{r}^{j}$ and $\mathbf{y}^{j}$ are the real labels of reconstructed image $\mathbf{\bar{X}}_{i,j}$ and image $\mathbf{{X}}_{i,j}$, respectively, and
\begin{equation*}
\mathbf{{y}}_{p}^{j}=\mbox{D}({W}_{\mbox{D}_i},\mathbf{{X}}_{i,j}).    
\end{equation*}
Hence, $\mbox{D}_i(\cdot)$ predicts whether an image comes from the set of input images or the set of reconstructed images. 

The next step optimises Equation~\ref{eq:opt-g} and Equation~\ref{eq:dis_loss} based on the updated parameters from the current step. The parameters of the one-class classifiers are then secret-shared with the service provider and regulator. To this end, each user $u_i$ secret-shares with the service provider and regulator  \emph{only} the set of parameters of the trained reconstructor\footnote{$\mbox{D}_i$ is an auxiliary model for training that is not needed for prediction.}, ${W}_{\mbox{R}_i}^*$, using an additive secret sharing scheme.

Overall, in PrivEdge, the private training of a multi-class classifier is reduced to each user training the proposed one-class RAN locally, independent of other users and even the service provider. The main advantage of the proposed local one-class RAN is providing a faster and more accurate private training than that of encrypted or differentially private distributed learning methods. In fact, with the proposed local training, users employ their data in the clear using fast computational resources (e.g.~GPUs), without communicating to others or adding noise to the parameters.

\section{Private prediction}
\label{sec:prediction}

The goal of the prediction phase is to determine whether a test image, $\mathbf{X} \in \mathbb{Z}_q^{w \times h \times c}$ ($q=2^8$ for images), held by user $C$ belongs to any of the $N$ classes or none of them. We aim to execute this classification while keeping $\mathbf{X}$  as well as the parameters of the $N$ one-class classifiers private. To this end, we secret-share $\mathbf{X}$ with the service provider and the regulator, and perform the \textit{private reconstruction} of each one-class classifier followed by \textit{dissimilarity based prediction}. 

\begin{figure}[t]
\centering
   \includegraphics[width=0.85\columnwidth]{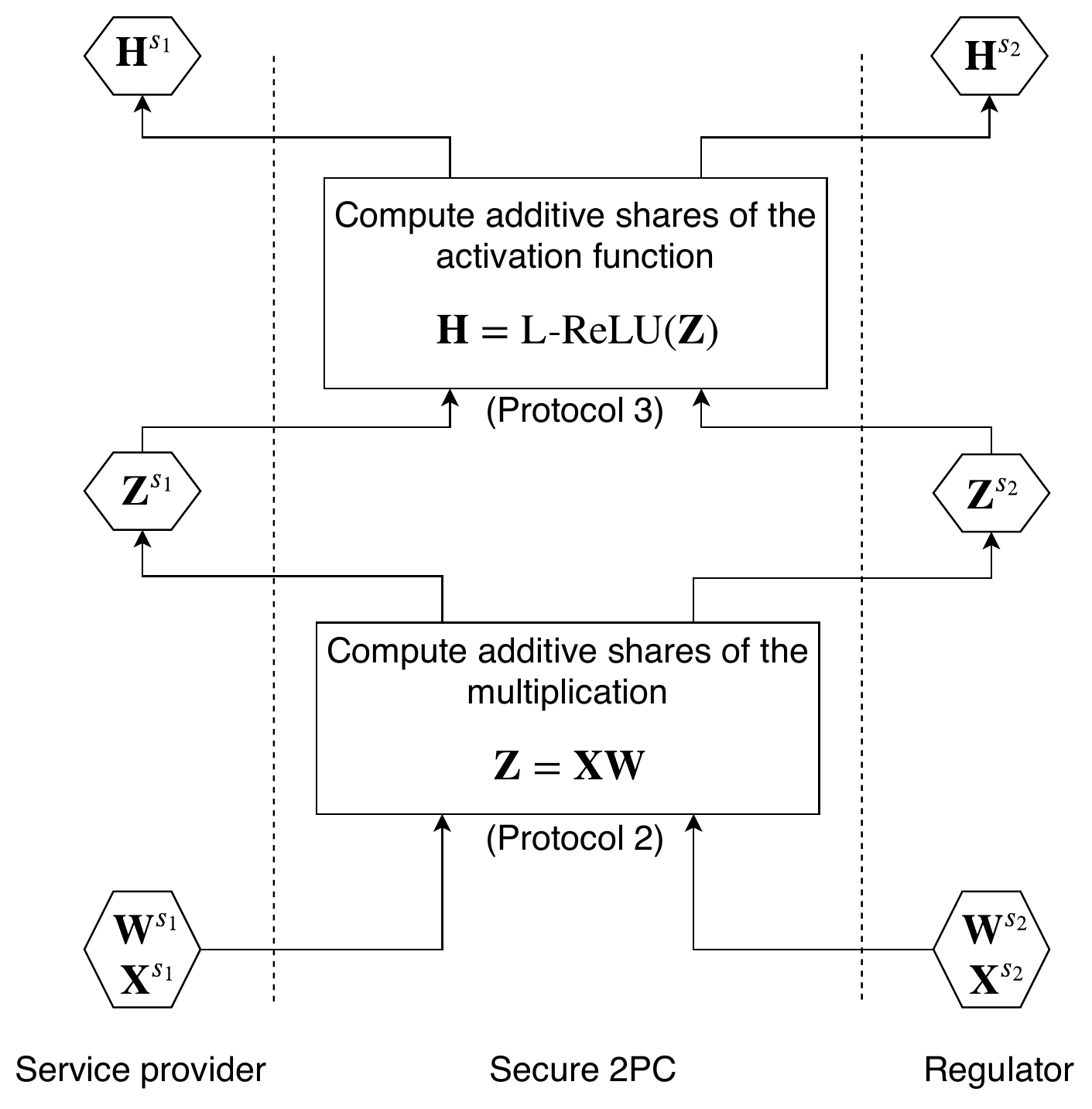}
  \caption{Our secure 2PC protocol for privately evaluating each layer of the one-class reconstructor. $\mathbf{W}^{\mbox{s}_1}$ and $\mathbf{W}^{\mbox{s}_2}$ are secret shares of the parameters and $\mathbf{X}^{\mbox{s}_1}$ and $\mathbf{X}^{\mbox{s}_2}$ are secret shares of the test image.}
  \label{fig:OneLayerRAN}
\end{figure}

First, we describe the MPC protocols that the service provider and regulator use to securely perform the multiplication and non-linear L-ReLU activation function of the reconstruction (see Figure~\ref{fig:OneLayerRAN}). Then, we describe the prediction based on the reconstruction dissimilarity.

\subsection{Private reconstruction}
\label{sec:P-Rec}

Recall that each $\mbox{R}_i(\cdot)$ consists of $T$ layers, where each layer is a linear transformation (i.e. matrix multiplications and additions) followed by a nonlinear transformation (i.e. activation function). To reduce the complexity of the prediction with 2PC, instead of using pooling layers, we use a convolutional stride larger than one.

\subsubsection{Input and parameters sharing}
To blind the parameters of all $T$ layers of $\mbox{R}_i(\cdot)$, each user $\mbox{u}_i$ runs \textbf{Protocol~\ref{prot:AddtiveSecret}} $T$ times on their set of parameters, ${W}_{{R}_i}^*$. By running \textbf{Protocol~\ref{prot:AddtiveSecret}}, each user $u_i$ randomly generates a set of matrices with the same size as the set of parameters 
\begin{equation*}  
    R = \{{\mathbf{R}}^{t} \in \mathbb{Z}_q^{w_k \times h_k \times c_t \times c_{t+1}}: t =1,..,T \},
\end{equation*}
which are sent to the service provider, $\mbox{s}_1$,
 \begin{equation*}
     {W}_i^{\mbox{s}_1}={R},
 \end{equation*}
  and to the regulator, $\mbox{s}_2$,
 \begin{equation*}
    {W}_i^{\mbox{s}_2}={W}_{{R}_i}^*-{R}.
 \end{equation*}
 
 Accordingly, $\mbox{s}_1$ and $\mbox{s}_2$ learn nothing about the parameters ${W}_{{R}_i}^*$ of each user $u_i$, as no information about the actual parameters can be gained from the individual shares in isolation. Moreover, as $\mathbf{X}$ may be a private image of one of the users, we additively secret-share it among $\mbox{s}_1$ and $\mbox{s}_2$ by running \textbf{Protocol~\ref{prot:AddtiveSecret}}:
\begin{equation*}
\label{eq:blind_inference_img}
\centering
\mathbf{X}^{\mbox{s}_1} = \mathbf{R}_x, \quad
\mathbf{X}^{\mbox{s}_2} = \mathbf{X} - \mathbf{R}_x,
\end{equation*}
where random $\mathbf{R}_x \in \mathbb{Z}_q^{w \times h \times c}$.

To reconstruct the image, $\mbox{s}_1$ and $\mbox{s}_2$ receive $\mathbf{X}^{\mbox{s}_1} \in \mathbb{Z}^{w \times h \times c}$ and $\mathbf{X}^{\mbox{s}_2} \in \mathbb{Z}^{w \times h \times c}$ and
 compute $\mathbf{\bar{X}}_i=\mbox{R}_i({W}_{{R}_i}^*,\mathbf{X})$ in a secure 2PC.

\subsubsection{Private linear transformation}
$\mbox{s}_1$ and $\mbox{s}_2$ perform the linear transformation of each layer by running \textbf{Protocol~\ref{prot:mult}}. \textbf{Protocol~\ref{prot:mult}} takes advantage of Beaver's precomputed multiplication technique, similarly to SecureML~\cite{mohassel2017secureml}, in which online time of multiplying additively secret shares can be improved by considering a data independent offline pre-computation. More concretely, to compute each multiplication of the $t$-th layer $z_t= (x_t^{\mbox{s}_1}+x_t^{\mbox{s}_2})(w_t^{\mbox{s}_1}+w_t^{\mbox{s}_2})$, we assume that two random values, $u_t$ and $v_t$ (and their product $q_t=u_tv_t$), are secret-shared among $\mbox{s}_1$ and $\mbox{s}_2$ during an offline pre-computation phase. $\mbox{s}_1$ and $\mbox{s}_2$ mask their shares by locally computing
\begin{protocol}[t]
\DontPrintSemicolon
\caption{Additive secret sharing}\label{prot:AddtiveSecret}
        \KwIn{a secret $\mathbf{I} \in \mathbb{Z}_q^{w \times h}$ ($q$ is $2^8$) }
        \KwOut{Two additive secret shares $\mathbf{S}_1 \in \mathbb{Z}_q^{w \times h}$ and $\mathbf{S}_2 \in \mathbb{Z}_q^{w \times h}$}
\BlankLine
\begin{minipage}{\hsize}
  \begin{algorithmic}[1]
    \STATE Generate random $\mathbf{R} \in \mathbb{Z}_q^{w \times h}$.
    \STATE Set secret shares\\
    $\mathbf{S}_1 = \mathbf{R}$ and $\mathbf{S}_2 = \mathbf{I} - \mathbf{R}$.
\end{algorithmic}
\end{minipage}
\end{protocol}
\begin{protocol}[t]
\DontPrintSemicolon
\caption{Multiplication}\label{prot:mult}
        {\bf Parties:} Service provider, $\mbox{s}_1$, and regulator, $\mbox{s}_2$\\
        \KwIn{Additive shares $\mathbf{X}^{s_1},\mathbf{X}^{s_2}$ and additive shares $\mathbf{W}^{\mbox{s}_1},\mathbf{W}^{\mbox{s}_2}$ and additive shares of multiplication triplets $\mathbf{Q}=\mathbf{U}\mathbf{V}$ }
        \KwOut{Additive shares of $\mathbf{Z} = (\mathbf{X}^{s_1}+\mathbf{X}^{s_2})(\mathbf{W}^{\mbox{s}_1}+\mathbf{W}^{\mbox{s}_2})$}
\BlankLine
\begin{minipage}{\hsize}
  \begin{algorithmic}[1]
    \STATE $\mbox{s}_1$ masks its shares using $\mathbf{U}^{s_1}$ and $\mathbf{V}^{s_1}$: \\
    $\mathbf{E}^{s_1}=\mathbf{X}^{s_1}-\mathbf{U}^{s_1}, \mathbf{F}^{s_1}=\mathbf{W}^{s_1}-\mathbf{V}^{s_1}$.
    \STATE $\mbox{s}_2$ masks its shares using $\mathbf{U}^{s_2}$ and $\mathbf{V}^{s_2}$: \\
    $\mathbf{E}^{s_2}=\mathbf{X}^{s_2}-\mathbf{U}^{s_2}, \mathbf{F}^{s_2}=\mathbf{W}^{s_2}-\mathbf{V}^{s_2}$.
    \STATE $\mbox{s}_1$ recovers $\mathbf{E}$ and $\mathbf{F}$ by receiving 
    $\mathbf{E}^{s_2}$ and $\mathbf{F}^{s_2}$ from $\mbox{s}_2$:\\
    $\mathbf{E}=\mathbf{E}^{s_1}+\mathbf{E}^{s_2}, \mathbf{F}=\mathbf{F}^{s_1}+\mathbf{F}^{s_2}$.
    \STATE $\mbox{s}_2$ recovers $\mathbf{E}$ and $\mathbf{F}$ by receiving 
    $\mathbf{E}^{s_1}$ and $\mathbf{F}^{s_1}$ from $\mbox{s}_1$:\\
    $\mathbf{E}=\mathbf{E}^{s_2}+\mathbf{E}^{s_1}, \mathbf{F}=\mathbf{F}^{s_2}+\mathbf{F}^{s_1}$.
    \STATE $\mbox{s}_1$ sets  $\mathbf{Z}^{s_1}=\mathbf{F}\mathbf{X}^{s_1}+\mathbf{E}\mathbf{W}^{s_1}+\mathbf{Q}^{s_1}$.
    \STATE $\mbox{s}_2$ sets $\mathbf{Z}^{s_2}=-\mathbf{E}\mathbf{F}+\mathbf{F}\mathbf{X}^{s_2}+\mathbf{E}\mathbf{W}^{s_2}+\mathbf{Q}^{s_2}$.
\end{algorithmic}
\end{minipage}
\end{protocol}
\begin{equation*}
e_t^{s_1}=x_t^{\mbox{s}_1}-u_t^{\mbox{s}_1}, f_t^{\mbox{s}_1}=w_t^{\mbox{s}_1}-v_t^{\mbox{s}_1},  
\end{equation*}
and
\begin{equation*}
e_t^{\mbox{s}_2}=x_t^{\mbox{s}_2}-u_t^{\mbox{s}_2}, f_t^{\mbox{s}_2}=w_t^{\mbox{s}_2}-v_t^{\mbox{s}_2},
\end{equation*}
respectively. Then $\mbox{s}_1$ sends $e_t^{\mbox{s}_1}$ and $f_t^{\mbox{s}_1}$ to $\mbox{s}_2$, and $\mbox{s}_2$ sends $e_t^{\mbox{s}_2}$ and $f_t^{\mbox{s}_2}$ to $\mbox{s}_1$, and each locally reconstructs $e_t$ and $f_t$ by adding $e_t^{\mbox{s}_1}$, $e_t^{\mbox{s}_2}$ and $f_t^{\mbox{s}_1}$, $f_t^{\mbox{s}_2}$, respectively. Next, $\mbox{s}_1$ computes
\begin{equation*}
    z_t^{\mbox{s}_1}=f_tx_t^{\mbox{s}_1}+e_tw_t^{\mbox{s}_1}+q_t^{\mbox{s}_1},
\end{equation*}
and $\mbox{s}_2$ computes
\begin{equation*}
z_t^{\mbox{s}_2}=-e_tf_t+f_tx_t^{\mbox{s}_2}+e_tw_t^{\mbox{s}_2}+q_t^{\mbox{s}_2}.
\end{equation*}

At the end of these interactions, $\mbox{s}_1$ and $\mbox{s}_2$ hold $\mathbf{Z}_t^{\mbox{s}_1}$ and $\mathbf{Z}_t^{\mbox{s}_2}$, additive shares of the linear transformation's output, $\mathbf{Z}_t$.

\subsubsection{Private non-linear function}
To compute the L-ReLU activation function securely on each element $z$ of $\mathbf{z}$:
\begin{equation}
  \mbox{L-ReLU}(z)=\begin{cases}
    z & \text{if $z>0$}\\
    \alpha z & \text{otherwise}
  \end{cases}
  \label{eq:lrelu}
\end{equation}
and obtain a secret share of the output of each layer $h$, $\mbox{s}_1$ and $\mbox{s}_2$ run \textbf{Protocol~\ref{prot:L-ReLU}}, a Yao's Garbled Circuit protocol~\cite{Yao:1986:GES:1382439.1382944} that securely performs an L-ReLU activation function by representing it as a Boolean circuit, $C$. Our proposed functionality of $C$ includes a 2-input multiplexer, which chooses among the reconstructed $z$ and $\alpha z$\footnote{To construct shares of $\alpha z$ the parties only need to multiply their shares by the public values $\alpha$ {\em locally}.} based on the first bit of $z$, denoted $z[1]$\footnote{As we use the common two's complement encoding, $z[1]$ corresponds to the sign $z$.}. Circuit $C$ selects $z$ if $z[1]$ is zero, otherwise $\alpha z$. Finally, we perform another addition to obtain the secret shares of the following layer:
\begin{protocol}[t]
\DontPrintSemicolon
\caption{L-ReLU}\label{prot:L-ReLU}
        {\bf Parties:} Service provider, $\mbox{s}_1$, and regulator, $\mbox{s}_2$\\
        \KwIn{Additive shares $\mathbf{Z}^{s_1}$ and $\mathbf{Z}^{s_2}$, and random $\mathbf{R'}$ }
        \KwOut{Additive shares of $\mathbf{H} = \mbox{L-ReLU}(\mathbf{Z}^{s_1}+\mathbf{Z}^{s_2})$}
\BlankLine
\begin{minipage}{\hsize}
  \begin{algorithmic}[1]
    \STATE $\mbox{s}_1$ and $\mbox{s}_2$ create Boolean circuit $C$ of L-ReLU followed by an addition.
    \STATE $\mbox{s}_1$ garbles $C$.
    \STATE $\mbox{s}_1$ sends its garbled inputs, $C$ and truth table to $\mbox{s}_2$.
    \STATE $\mbox{s}_2$ receives garbled values correspondence to its input via OT.
    \STATE $\mbox{s}_2$ evaluates the garbled $C$ using its garbled input and $\mbox{s}_1$'s garbled input and compute garbled $\mathbf{H}-\mathbf{R'}$.
    \STATE $\mbox{s}_1$ sets $\mathbf{H}^{s_1}=\mathbf{R'}$.
    \STATE $\mbox{s}_2$ sets $\mathbf{H}^{s_2}=\mathbf{H}-\mathbf{R'}$.
\end{algorithmic}
\end{minipage}
\end{protocol}
\begin{equation*}
\centering
\mathbf{X}_{t+1}^{\mbox{s}_1} = \mathbf{R}', \quad
\mathbf{X}_{t+1}^{\mbox{s}_2} = \mathbf{H}_t - \mathbf{R}',
\end{equation*}
which are the inputs of $\mbox{s}_1$ and $\mbox{s}_2$ for the next layer. This process is repeated for all the layers of each $\mbox{R}_i(\cdot)$.      

To perform the mentioned computation inside $C$ securely, as a {\em garbler}, $\mbox{s}_1$ garbles $C$ gate-by-gate. By garbling we mean that $\mbox{s}_1$ generates two random labels for each wire of $C$ and encrypts the output labels of each gate by using the corresponding input labels as encryption keys. Then, $\mbox{s}_1$ sends to $\mbox{s}_2$  the garbled $C$ with garbled truth tables, which show all possible garbled values that each gate can attain.

As an {\em evaluator}, $\mbox{s}_2$ computes garbled $C$ gate-by-gate from the input wires to the output wires. $\mbox{s}_2$ needs to know the labels of their input and the $\mbox{s}_1$ input prior to the computation. $\mbox{s}_1$ sends their input labels to $\mbox{s}_2$ directly, as $\mbox{s}_2$ cannot learn the real input of $\mbox{s}_1$ from its random label. However, $\mbox{s}_2$ asks the label of its input via Oblivious Transfer (OT)~\cite{goldreichbook}, which guarantees that $\mbox{s}_2$ only receives the label corresponding to their input from $\mbox{s}_1$, while $\mbox{s}_1$ learns nothing from the input of $\mbox{s}_2$. In addition, neither $\mbox{s}_1$ nor $\mbox{s}_2$ learns intermediate values in the clear. We refer the reader to~\cite{SEC-019} for a detailed presentation and security analysis of Garbled Circuits.

\subsection{Dissimilarity based prediction}
\label{sec:DissPred}

\begin{figure}[t]
 \centering
  \includegraphics[width=0.5\textwidth]{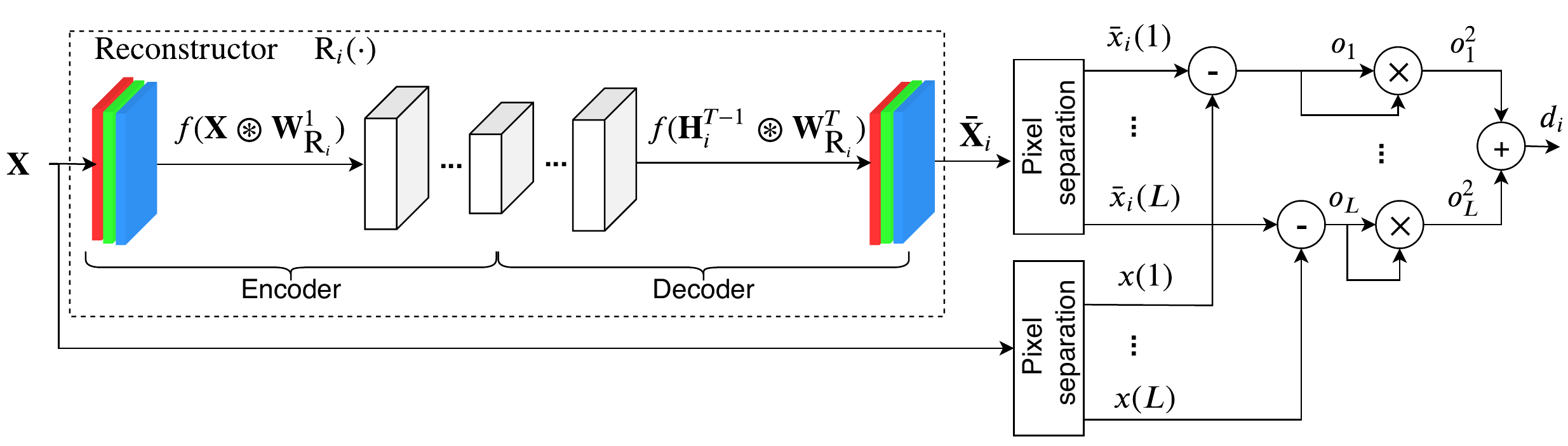}
  \caption{ Computation of the reconstruction dissimilarity of each one-class reconstructor $\mbox{R}_i(\cdot)$ with $T$ layers in PrivEdge given an image $\mathbf{X}$ with $L=w \times h\times c$ pixels, where $w$, $h$ and $c$ are width, height and number of channels of the image, respectively. The uploaded image $\mathbf{X}$ is passed through all $T$ layers of $\mbox{R}_i(\cdot)$ to produce the reconstructed image, $\mathbf{\bar{X}}_i$. The output of each first layer, $\mathbf{H}_i^1$, is computed as the convolution $\circledast$ of the input of this layer, $\mathbf{X}$, and its corresponding parameters, $\mathbf{W}_{\mbox{R}_i}^1$, followed by L-ReLU activation function $f(\cdot)$. Finally, the reconstruction dissimilarity, $d_i$, between the uploaded image and the reconstructed image is measured as the sum square difference in each first pixel, $o_1 = x(1) - \bar{x}_i(1)$, for example. }
  \label{fig:global_filter}
\end{figure}

$\mbox{s}_1$ and $\mbox{s}_2$ hold secret shares of the input image, $\mathbf{X}$, and the reconstructed image, $\mathbf{\bar{X}}$. To classify  $\mathbf{X}$, $\mbox{s}_1$ and $\mbox{s}_2$ compute the reconstruction dissimilarity (see Figure~\ref{fig:global_filter}), $d_i$, between $\mathbf{X}$ and $\mathbf{\bar{X}}_i$ (for each $i=1, ...,N$) and assign the image $\mathbf{X}$ to the class corresponding to the minimum reconstruction dissimilarity  (or none of them if the smallest $d_i$ is above a predefined secret-shared threshold) by securely computing subtraction and multiplication in the Arithmetic circuit followed by computing the minimum and comparison via Yao's Garbled Circuit protocol. In a social media application, if the assigned class differs from the class of the user who uploaded image $\mathbf{X}$ {\emph{and}} the reconstruction dissimilarity is smaller than a {\em privacy threshold}, then the image is blocked (i.e.~not shared). In Section~\ref{sec:result} we evaluate the effect of the privacy threshold with a specific dissimilarity function.

\begin{figure}[t]
\centering
   \includegraphics[height=0.45\textheight, width=\columnwidth]{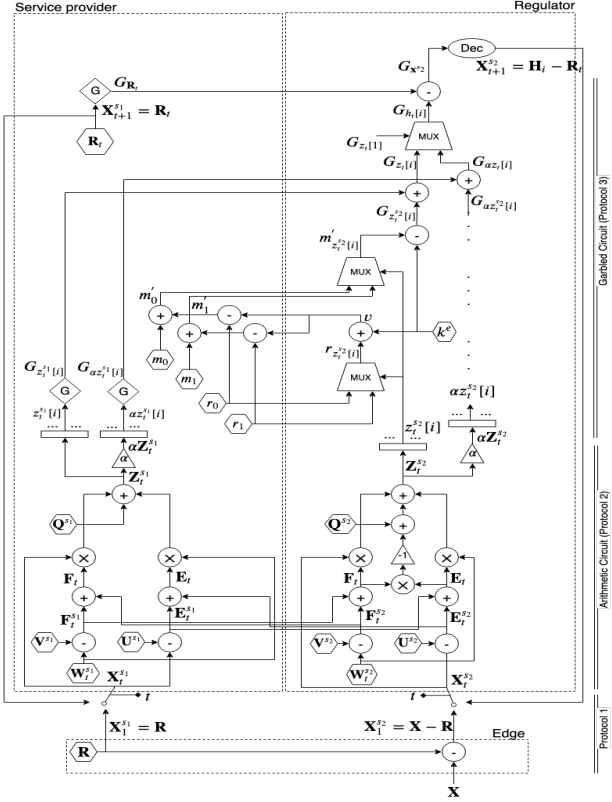}
  \caption{The private prediction of each reconstructor by the service provider $\mbox{s}_1$ and regulator $\mbox{s}_2$ for an image $\mathbf{X}$.
  Each layer $t$ is a multiplication between the input, $\mathbf{X}_t$, and the parameters, $\mathbf{W}_t$, of this layer, $\mathbf{Z}_t=\mathbf{X}_t \mathbf{W}_t$, followed by a L-ReLU activation function, $\mathbf{H}_t=\text{L-ReLU}(\mathbf{Z}_t)$. The parameters and inputs of each layer $t$ are secret-shared among $\mbox{s}_1$ and $\mbox{s}_2$ as $\mathbf{X}_t^{\mbox{s}_1}$, $\mathbf{X}_t^{\mbox{s}_2}$ and $\mathbf{W}_t^{\mbox{s}_1}$, $\mathbf{W}_t^{\mbox{s}_2}$, respectively, to preserve their privacy. The input to the first layer of the reconstructor is the user's data, $\mathbf{X}$, which is masked using a random matrix $\mathbf{R}$ to be secret-shared among $\mbox{s}_1$ and $\mbox{s}_2$ (\textbf{Protocol~\ref{prot:AddtiveSecret}}). To do the private multiplication of each layer $t$, $\mbox{s}_1$ and $\mbox{s}_2$ locally mask their input and parameters using their shares of random matrices $\mathbf{U}$ and $\mathbf{V}$, and exchange them to obtain their additive shares, $\mathbf{Z}_t^{\mbox{s}_1}$ and $\mathbf{Z}_t^{\mbox{s}_2}$, and multiply them by the public value $\alpha$ locally (\textbf{Protocol~\ref{prot:mult}}). $\mbox{s}_1$ and $\mbox{s}_2$ engage in \textbf{Protocol~\ref{prot:L-ReLU}} to privately perform the L-ReLU activation function, which is represented using a Boolean circuit. This Boolean circuit includes a 2-input multiplexer (MUX) that outputs $\mathbf{Z}_t$ if its first bit, $z[1]$, is zero, otherwise $\alpha \mathbf{Z}_t$. $\mbox{s}_1$ sends the garbled values of their inputs $\mathbf{G}$ via Oblivious transfer to the evaluator of the garbled circuit, $\mbox{s}_2$. In Oblivious transfer, for each $i$-th bit of the secret share of $\mbox{s}_2$, $z^{\mbox{s}_2}[i]=\{0,1\}$, $\mbox{s}_2$ receives the garbled value $G_{z^{\mbox{s}_2}[i]}$ which equals to one of two garbled messages $m_0$ and $m_1$ of $\mbox{s}_1$ based on the value of $z^{\mbox{s}_2}[i]$. $\mbox{s}_1$ generates two random values ($r_0$ and $r_1$) and RSA keys (the modulus $N$ and the public exponent $e$) and sends them to $\mbox{s}_2$. $\mbox{s}_2$ chooses one of the random values based on the value of their input and masks it by generating a random value $k$, $v=(r_b+k^e) \mbox{mod} N$. Then, $\mbox{s}_1$ receives $v$ and masks their messages $m'_0=m_0+k_0$ and $m'_1=m_1+k_1$, where $k_0=(v-x_0 )^d \mbox{mod} N$ and $k_1=(v-x_1 )^d \mbox{mod} N$. One of $k_0$ or $k_1$ is equal to $k$.  $\mbox{s}_2$ receives the masked messages and unmasks the corresponding message, $G_{z^{\mbox{s}_2}[i]}=m'_{z^{\mbox{s}_2}[i]}-k$. We also consider another addition after computing L-ReLU in the garbled circuit to obtain the secret shares of the input of the next layer. }
  \label{fig:inference_RAN}
\end{figure}

Overall, the service provider and the regulator together perform the private prediction of each one-class classifier using different 2PC protocols for different operations to achieve substantial gains in complexity and speed (see Figure~\ref{fig:inference_RAN}). In the choices of 2PC \textbf{Protocols \ref{prot:AddtiveSecret}, \ref{prot:mult}} and~\textbf{\ref{prot:L-ReLU}}, we leverage the fact that non-linear transformations such as L-ReLU can be represented efficiently as simple Boolean circuits, which are best computed using Yao's Garbled Circuit~\cite{juvekar2018gazelle}, while linear transformations such as addition and multiplication have an efficient representation as secure arithmetic circuits~\cite{mohassel2017secureml}.


\section{Validation}
\label{sec:result}

In this section, we discuss the evaluation of PrivEdge for privacy-preserving MLaaS. We evaluate PrivEdge in recognising users from their faces, handwritten text and letters.

\subsection{Datasets and Architecture}
We consider three realistic privacy-sensitive scenarios with the Internet Movie Data Base (IMDB) dataset~\cite{parkhi2015deep}, Informatics and Applied Mathematics handwritten (IAM) dataset~\cite{marti2002iam} and Russian handwritten lowercase letters dataset~\cite{BaselineRussianLetter}. Here, we introduce the datasets and state-of-the-art multi-class classifiers that are trained directly on them in a centralised manner without considering privacy.

We adopted a subset of celebrities from the \emph{IMDB} dataset~\cite{parkhi2015deep} as users. IMDB includes whole body images with variations between images of each class, such as the number of individuals within images, and their rotation, pose, and illumination. We detect and crop the faces~\cite{wolf2011face} to obtain $128 \times 128 \times 3 $ images. As an $N$-class classifier, we use 7 convolutional layers followed by 2 fully-connected layers and a softmax layer 32C4-64C4-128C4-256C4-256C4-256C4-256C4-256FC-256FC-$N$SM, where 32C5 is a convolutional layer with 32 kernels of size $4\times4$, 256FC is a fully-connected layer with size 256 and $N$SM is a softmax layer of size $N$ (the number of classes).

The \emph{IAM handwritten} dataset~\cite{marti2002iam} contains scanned pages of handwritten English text. Each user $u_i$ collects their greyscale pages of scanned text and break down the greyscale lines to image size ($128 \times 128 \times 1$) without considering breaking them with respect to sentences or words. As the width of the lines are larger than their heights, and to keep the aspect ratio during resizing, we first resize the height of the text to $128$, followed by resizing the width with the same height factor. Then, we crop the text with size $128 \times 128$ with random starting point of width and select 0.1 of them. As proposed in~\cite{BaselineIAM}, we use a convolutional 10-class classifier 32C5-MP-64C3-MP-128C3-MP-DO-512FC-DO-256FC-DO-10SM, where MP and DO stand for Max Pooling and Drop Out.

\emph{Russian lowercase letters} includes 33 classes of handwritten Russian lowercase letters $(32 \times 32 \times 3 )$. The width and height of the letters are smaller than the IMDB and IAM samples, and so we use a shallower one-class classifier. As proposed in~\cite{BaselineRussianLetter}, we use convolutional 33-class classifier 32C3-32C3-DO-256FC-33SM.

\subsection{PrivEdge: implementation details}

In contrast to training an $N$-class classifier on a centralised dataset, users train one-class RANs ($T=14$) locally on their private data (i.e. faces, handwritten texts or handwritten letters) using a mini-batch Adam optimizer~\cite{kingma2014adam} with $\text{learning rate}=0.0002$, $\gamma=0.999$, and $\beta=0.001$. The reconstructor encoder and decoder both have seven layers for IMDB and IAM, and six for the Russian lowercase letters dataset. The discriminator of the RAN classifies data against the reconstructed data through 5 convolutional layers, with $w_k=h_k=4$ (width and height of the kernel). L-ReLU is the chosen activation function. The stride sizes of the encoder, decoder and discriminator are 2, 1 and 2 respectively. The training phases of the one-class RANs were implemented in Python with the publicly available Keras library~\cite{chollet2015keras} on a remote server with one NVIDIA Tesla P100 GPU. The training uses several iterations, each with 32 randomly selected training data samples. This random selection enables the model to see all of the training images, while converging more quickly than when using epochs and iterating through the entire dataset within each epoch~\cite{kerasgan}. We used the ABY~\cite{demmler2015aby} library for secure 2PC (i.e. additive secret-sharing and Garbled circuit) with 128-bit security parameter and SIMD circuits running the service provider and the regulator on an Intel(R) Xeon(R) CPU E5-2690 v3 @ 2.60GHz parallel with 24 cores.

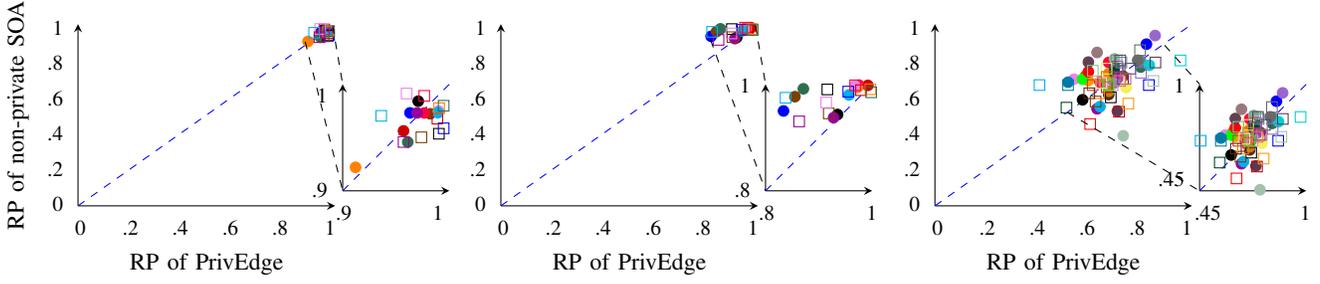
\begin{figure*}
          \begin{tikzpicture}
          \begin{axis}[
                  name =ax1,
                  small,
                  axis lines=left, 
                  xtick={.0, .2, .4, .6, .8, 1},
                  xticklabels={0, .2, .4, .6, .8, 1},
                  ytick={.0, .2, .4, .6, .8, 1},
                  yticklabels={0, .2, .4, .6, .8, 1},
                  width=5cm,
                  height=4cm,
                  cycle list name=color list,
                  xmin=0.00,
                  ymin=0.00,
                  ymax=1.01,
                  xmax=1.01,
                  smooth,
                  ylabel= RP of non-private SOA,
                  enlarge y limits=0.01,
                  enlarge x limits=0.01,
                  xlabel={RP of PrivEdge}
                  ]
\addplot[
    scatter/classes={b={blue}, r={red}, g={green}, d={black}, c={chocolate(traditional)}, e={bostonuniversityred}, a={americanrose}, o={orange}, dg={darkgreen}, vw={violet(web)}, v={viola}, cy={cyan} , db={darkblue}, gi={giallo}, ro={rosso}  },
    scatter,
    only marks,
    scatter src=explicit symbolic,
    ] table [meta=Class] {
         x          y     Class     
    0.9712389	0.98474944	d
    0.9633621	0.97356826	b
    0.95717883	0.9563107	e
    0.982699	0.9726776	c
    0.9770115	0.973466	a
    0.91129035	0.9214876	o
    0.961206	0.9456265	dg
    0.99071616	0.9753247	vw
    0.97010463	0.9735294	v
    0.99006623	0.9737249	cy
};\label{fig:Recall_IAMvsB}

\addplot[
    scatter/classes={ b={blue}, r={red}, g={green}, d={black}, c={chocolate(traditional)}, e={bostonuniversityred}, a={americanrose}, o={orange}, dg={darkgreen}, vw={violet(web)}, v={viola}, cy={cyan} , db={darkblue}, gi={giallo}, ro={rosso}  },
    scatter,
    only marks,
    mark=square,
    scatter src=explicit symbolic,
    ] table [meta=Class] {
    x           y     Class       
0.9909707	0.953586	d
0.9955457	0.95878524	b
0.9895833	0.96805894	e
0.974271	0.9501779	c
0.9770115	0.98988193	a
0.99122804	0.9780702	o
0.99553573	0.98039216	dg
0.96015424	0.99207395	vw
0.9572271	0.9457143	v
0.93583727	0.97072417	cy
};\label{fig:Precision_ImdbvsB}
\coordinate (c11) at (axis cs:0.9, 0.92);
\coordinate (c22) at (axis cs:1.01, 1.01);
\draw [blue,dashed] (rel axis cs:0,0) -- (rel axis cs:1,1);

\end{axis}

\begin{axis}[
                  small,
                  name=ax2,
                  axis lines=left, 
                  xtick={0.9, .99},
                  ytick={0.9, .99},
                 yticklabels={.9,1},
                 xticklabels={.9,1},
                  width=3cm,
                  height=3cm,
                  cycle list name=color list,
                  xmax=1,
                  ymax=1,
                  xmin=0.9,
                  ymin=0.9,
                  smooth,
                  at={($(ax1.south east)+(0.1cm,0.2cm)$)},
                  clip=true,
                  enlarge y limits=0.01,
                  enlarge x limits=0.01,
                  ]
\addplot[
    scatter/classes={b={blue}, r={red}, g={green}, d={black}, c={chocolate(traditional)}, e={bostonuniversityred}, a={americanrose}, o={orange}, dg={darkgreen}, vw={violet(web)}, v={viola}, cy={cyan} , db={darkblue}, gi={giallo}, ro={rosso}  },
    scatter,
    only marks,
    scatter src=explicit symbolic,
    ] table [meta=Class] {
       x          y     Class     
    0.9712389	0.98474944	d
    0.9633621	0.97356826	b
    0.95717883	0.9563107	e
    0.982699	0.9726776	c
    0.9770115	0.973466	a
    0.91129035	0.9214876	o
    0.961206	0.9456265	dg
    0.99071616	0.9753247	vw
    0.97010463	0.9735294	v
    0.99006623	0.9737249	cy
};\label{fig:Recall_ImdbvsB}

\addplot[
    scatter/classes={ b={blue}, r={red}, g={green}, d={black}, c={chocolate(traditional)}, e={bostonuniversityred}, a={americanrose}, o={orange}, dg={darkgreen}, vw={violet(web)}, v={viola}, cy={cyan} , db={darkblue}, gi={giallo}, ro={rosso}  },
    scatter,
    only marks,
    mark=square,
    scatter src=explicit symbolic,
    ] table [meta=Class] {
    x           y     Class       
0.9909707	0.953586	d
0.9955457	0.95878524	b
0.9895833	0.96805894	e
0.974271	0.9501779	c
0.9770115	0.98988193	a
0.99122804	0.9780702	o
0.99553573	0.98039216	dg
0.96015424	0.99207395	vw
0.9572271	0.9457143	v
0.93583727	0.97072417	cy
};\label{fig:Precision_ImdbvsB}

\draw [blue,dashed] (rel axis cs:0,0) -- (rel axis cs:1,1);

\end{axis}
\draw [dashed] (c11) -- (ax2.south west);
\draw [dashed] (c22) -- (ax2.north west);
\end{tikzpicture}
          \begin{tikzpicture}
          \begin{axis}[
                  name =ax1,
                  small,
                  axis lines=left, 
                  xtick={.0, .2, .4, .6, .8, 1},
                  xticklabels={0, .2, .4, .6, .8, 1},
                  ytick={.0, .2, .4, .6, .8, 1},
                  yticklabels={0, .2, .4, .6, .8, 1},
                  width=5cm,
                  height=4cm,
                  cycle list name=color list,
                  xmin=0.00,
                  ymin=0.00,
                  ymax=1.01,
                  xmax=1.01,
                  smooth,
                  enlarge y limits=0.01,
                  enlarge x limits=0.01,
                  xlabel={RP of PrivEdge}
                  ]
\addplot[
    scatter/classes={b={blue}, r={red}, g={green}, d={black}, c={chocolate(traditional)}, e={bostonuniversityred}, a={americanrose}, o={orange}, dg={darkgreen}, vw={violet(web)}, v={viola}, cy={cyan} , db={darkblue}, gi={giallo}, ro={rosso}  },
    scatter,
    only marks,
    scatter src=explicit symbolic,
    ] table [meta=Class] {
   x             y            Class     

0.9348051     0.94377372      d   
0.83214285    0.95057143      b
0.9923034     0.99944025      e  
0.853821755   0.97780909      c 
0.975472895   0.99967938      a 
0.972504235   0.99576988      o 
0.86983617    0.99287749      dg 
0.96733083    0.99860982      vw 
0.927213915   0.93772746      v 
0.95644115    0.98123262      cy 
};\label{fig:Recall_IAMvsB}

\addplot[
    scatter/classes={ b={blue}, r={red}, g={green}, d={black}, c={chocolate(traditional)}, e={bostonuniversityred}, a={americanrose}, o={orange}, dg={darkgreen}, vw={violet(web)}, v={viola}, cy={cyan} , db={darkblue}, gi={giallo}, ro={rosso}  },
    scatter,
    only marks,
    mark=square,
    scatter src=explicit symbolic,
    ] table [meta=Class] {
   x             y            Class       
0.915385975     0.99166365     d
0.955089535   0.98753339       b
0.97928405      0.99029395     e    
0.918029995   0.94582624       c
0.9681831  0.99967938          a
0.99718022    0.99199326       o
0.99856836    0.98620446       dg
0.913185115     0.96680126     vw
0.86186963    0.93057785       v
0.8360682    0.97580645        cy
};\label{fig:Precision_IAMvsB}
\coordinate (c11) at (axis cs:0.83, 0.93);
\coordinate (c22) at (axis cs:1.01, 1.01);
\draw [blue,dashed] (rel axis cs:0,0) -- (rel axis cs:1,1);

\end{axis}

\begin{axis}[
                  small,
                  name=ax2,
                  axis lines=left, 
                  xtick={0.8, 1},
                  ytick={0.8, 1},
                  yticklabels={.8,1},
                  xticklabels={.8,1},
                  width=3cm,
                  height=3cm,
                  cycle list name=color list,
                  xmin=0.8,
                  ymin=0.8,
                  smooth,
                  at={($(ax1.south east)+(0.1cm,0.2cm)$)},
                  clip=true,
                  enlarge y limits=0.01,
                  enlarge x limits=0.01,
                  ]
\addplot[
    scatter/classes={b={blue}, r={red}, g={green}, d={black}, c={chocolate(traditional)}, e={bostonuniversityred}, a={americanrose}, o={orange}, dg={darkgreen}, vw={violet(web)}, v={viola}, cy={cyan} , db={darkblue}, gi={giallo}, ro={rosso}  },
    scatter,
    only marks,
    scatter src=explicit symbolic,
    ] table [meta=Class] {
   x             y            Class     

0.9348051     0.94377372      d   
0.83214285    0.95057143      b
0.9923034     0.99944025      e  
0.853821755   0.97780909      c 
0.975472895   0.99967938      a 
0.972504235   0.99576988      o 
0.86983617    0.99287749      dg 
0.96733083    0.99860982      vw 
0.927213915   0.93772746      v 
0.95644115    0.98123262      cy 
};\label{fig:Recall_IAMvsB}

\addplot[
    scatter/classes={ b={blue}, r={red}, g={green}, d={black}, c={chocolate(traditional)}, e={bostonuniversityred}, a={americanrose}, o={orange}, dg={darkgreen}, vw={violet(web)}, v={viola}, cy={cyan} , db={darkblue}, gi={giallo}, ro={rosso}  },
    scatter,
    only marks,
    mark=square,
    scatter src=explicit symbolic,
    ] table [meta=Class] {
   x             y            Class       
0.915385975     0.99166365     d
0.955089535   0.98753339       b
0.97928405      0.99029395     e    
0.918029995   0.94582624       c
0.9681831  0.99967938          a
0.99718022    0.99199326       o
0.99856836    0.98620446       dg
0.913185115     0.96680126     vw
0.86186963    0.93057785       v
0.8360682    0.97580645        cy
};\label{fig:Precision_IAMvsB}

\draw [blue,dashed] (rel axis cs:0,0) -- (rel axis cs:1,1);

\end{axis}
\draw [dashed] (c11) -- (ax2.south west);
\draw [dashed] (c22) -- (ax2.north west);

\end{tikzpicture}
\begin{tikzpicture}
          \begin{axis}[
                  small,
                  name=ax1,
                  axis lines=left, 
                  width=5cm,
                  height=4cm,
                  cycle list name=color list,
                  xtick={.0, .2, .4, .6, .8, 1},
                  xticklabels={0, .2, .4, .6, .8, 1},
                  ytick={.0, .2, .4, .6, .8, 1},
                  yticklabels={0, .2, .4, .6, .8, 1},
                  xmax=1.01,
                  xmin=0.0,
                  ymin=0.0,
                  ymax=1.01,
                  smooth,
                  enlarge y limits=0.01,
                  enlarge x limits=0.01,
                  xlabel={RP of PrivEdge}
                  ]
\addplot[
    scatter/classes={b={blue}, r={red}, g={green}, d={black}, c={chocolate(traditional)}, e={bostonuniversityred}, a={americanrose}, o={orange}, dg={darkgreen}, vw={violet(web)}, v={viola}, cy={cyan} , db={darkblue}, gi={giallo}, ro={rosso}, af={airforceblue}, al={almond}, am={amethyst}, baz={bazaar}, br={britishracinggreen}, by={byzantine}, ca={cadetblue},  cam={cambridgeblue}, can={candypink}, cap={caputmortuum},  cer={cerulean}, co={corn}, da={darkbyzantium}, dag={darkgoldenrod}, das={darkseagreen}, dat={darkturquoise}, dim={dimgray}, eg={eggplant}},
    scatter,
    only marks,
    scatter src=explicit symbolic,
    ] table [meta=Class] {
   x           y      Class     
0.5813953	0.59090909	d
0.8372093	0.90625	    b
0.6511628	0.69230769	e
0.74418604	0.72463768	c
0.68604654	0.8028169	a
0.68604654	0.73972603	o
0.81395346	0.77647059	dg
0.54651165	0.70731707	vw
0.6395349	0.53982301	v
0.6511628	0.55238095	cy
0.59302324	0.725	    ro
0.60465115	0.70	    gi
0.7093023	0.77647059	db
0.5813953	0.70731707	g
0.60465115	0.75	    r
0.81395346	0.81176471	af
0.73255813	0.70238095	al
0.872093	0.95522388	am 
0.6395349	0.85915493	baz
0.7093023	0.74603175	br
0.7209302	0.71276596	by
0.7093023	0.68627451	ca
0.74418604	0.38709677	cam
0.6627907	0.68965517	can
0.68604654	0.62068966	cap
0.5232558	0.69047619	cer
0.75581396	0.66	    co
0.7209302	0.52671756	da
0.68604654	0.63917526	dag
0.7093023	0.82278481	das
0.8488372	0.78651685	dat
0.8023256	0.81521739	dim
0.60465115	0.8045977	eg
};\label{fig:Recall_RussianvsB}

\addplot[
    scatter/classes={b={blue}, r={red}, g={green}, d={black}, c={chocolate(traditional)}, e={bostonuniversityred}, a={americanrose}, o={orange}, dg={darkgreen}, vw={violet(web)}, v={viola}, cy={cyan} , db={darkblue}, gi={giallo}, ro={rosso}, af={airforceblue}, al={almond}, am={amethyst}, baz={bazaar}, br={britishracinggreen}, by={byzantine}, ca={cadetblue},  cam={cambridgeblue}, can={candypink}, cap={caputmortuum},  cer={cerulean}, co={corn}, da={darkbyzantium}, dag={darkgoldenrod}, das={darkseagreen}, dat={darkturquoise}, dim={dimgray}, eg={eggplant}},
    scatter,
    only marks,
    mark=square,
    scatter src=explicit symbolic,
    ] table [meta=Class] {
   x             y     Class       
0.6944444	0.60465116	d
0.84705883	0.6744186	b
0.72727275	0.52325581	e
0.63366336	0.58139535	c
0.6781609	0.6627907	a
0.64835167	0.62790698	o
0.72164947	0.76744186	dg
0.70149255	0.6744186	vw
0.7432432	0.70930233	v
0.40875912	0.6744186	cy
0.6219512	0.6744186	ro
0.7647059	0.56976744	gi
0.7721519	0.76744186	db
0.6849315	0.6744186	g
0.6117647	0.45348837	r
0.7368421	0.80232558	af
0.65625	    0.68604651	al
0.8152174	0.74418605	am
0.75342464	0.70930233	baz
0.5169492	0.54651163	br
0.7209302	0.77906977	by
0.7176471	0.81395349	ca
0.8648649	0.69767442	cam
0.7307692	0.69767442	can
0.6082474	0.62790698	cap
0.5232558	0.6744186	cer
0.6770833	0.76744186	co
0.87323946	0.80232558	da
0.6781609	0.72093023	dag
0.622449	0.75581395	das 
0.97333336	0.81395349	dat
0.8117647	0.87209302	dim
0.73239434	0.81395349  eg	
};\label{fig:Precision_RussianvsB}
\coordinate (c111) at (axis cs:0.52, 0.52);
\coordinate (c222) at (axis cs:0.91, 0.9);
\draw [blue,dashed] (rel axis cs:0,0) -- (rel axis cs:1,1);

\end{axis}

         \begin{axis}[
                  small,
                  name=ax2,
                  axis lines=left, 
                  width= 3cm,
                  height=3cm,
                  cycle list name=color list,
                  xmax=1,
                  xtick={0.45, 1},
                  ytick={0.45, 1},
                  yticklabels={.45,1},
                  xticklabels={.45,1},
                  ymax=1,
                  smooth,
                  at={($(ax1.south east)+(0.1cm,0.2cm)$)},
                  clip=true,
                  enlarge y limits=0.01,
                  enlarge x limits=0.01,
                  ]
\addplot[
    scatter/classes={b={blue}, r={red}, g={green}, d={black}, c={chocolate(traditional)}, e={bostonuniversityred}, a={americanrose}, o={orange}, dg={darkgreen}, vw={violet(web)}, v={viola}, cy={cyan} , db={darkblue}, gi={giallo}, ro={rosso}, af={airforceblue}, al={almond}, am={amethyst}, baz={bazaar}, br={britishracinggreen}, by={byzantine}, ca={cadetblue},  cam={cambridgeblue}, can={candypink}, cap={caputmortuum},  cer={cerulean}, co={corn}, da={darkbyzantium}, dag={darkgoldenrod}, das={darkseagreen}, dat={darkturquoise}, dim={dimgray}, eg={eggplant}},
    scatter,
    only marks,
    scatter src=explicit symbolic,
    ] table [meta=Class] {
   x           y      Class     
0.5813953	0.59090909	d
0.8372093	0.90625	    b
0.6511628	0.69230769	e
0.74418604	0.72463768	c
0.68604654	0.8028169	a
0.68604654	0.73972603	o
0.81395346	0.77647059	dg
0.54651165	0.70731707	vw
0.6395349	0.53982301	v
0.6511628	0.55238095	cy
0.59302324	0.725	    ro
0.60465115	0.70	    gi
0.7093023	0.77647059	db
0.5813953	0.70731707	g
0.60465115	0.75	    r
0.81395346	0.81176471	af
0.73255813	0.70238095	al
0.872093	0.95522388	am 
0.6395349	0.85915493	baz
0.7093023	0.74603175	br
0.7209302	0.71276596	by
0.7093023	0.68627451	ca
0.74418604	0.38709677	cam
0.6627907	0.68965517	can
0.68604654	0.62068966	cap
0.5232558	0.69047619	cer
0.75581396	0.66	    co
0.7209302	0.52671756	da
0.68604654	0.63917526	dag
0.7093023	0.82278481	das
0.8488372	0.78651685	dat
0.8023256	0.81521739	dim
0.60465115	0.8045977	eg
};\label{fig:Recall_RussianvsB}

\addplot[
    scatter/classes={b={blue}, r={red}, g={green}, d={black}, c={chocolate(traditional)}, e={bostonuniversityred}, a={americanrose}, o={orange}, dg={darkgreen}, vw={violet(web)}, v={viola}, cy={cyan} , db={darkblue}, gi={giallo}, ro={rosso}, af={airforceblue}, al={almond}, am={amethyst}, baz={bazaar}, br={britishracinggreen}, by={byzantine}, ca={cadetblue},  cam={cambridgeblue}, can={candypink}, cap={caputmortuum},  cer={cerulean}, co={corn}, da={darkbyzantium}, dag={darkgoldenrod}, das={darkseagreen}, dat={darkturquoise}, dim={dimgray}, eg={eggplant}},
    scatter,
    only marks,
    mark=square,
    scatter src=explicit symbolic,
    ] table [meta=Class] {
   x             y     Class       
0.6944444	0.60465116	d
0.84705883	0.6744186	b
0.72727275	0.52325581	e
0.63366336	0.58139535	c
0.6781609	0.6627907	a
0.64835167	0.62790698	o
0.72164947	0.76744186	dg
0.70149255	0.6744186	vw
0.7432432	0.70930233	v
0.40875912	0.6744186	cy
0.6219512	0.6744186	ro
0.7647059	0.56976744	gi
0.7721519	0.76744186	db
0.6849315	0.6744186	g
0.6117647	0.45348837	r
0.7368421	0.80232558	af
0.65625	    0.68604651	al
0.8152174	0.74418605	am
0.75342464	0.70930233	baz
0.5169492	0.54651163	br
0.7209302	0.77906977	by
0.7176471	0.81395349	ca
0.8648649	0.69767442	cam
0.7307692	0.69767442	can
0.6082474	0.62790698	cap
0.5232558	0.6744186	cer
0.6770833	0.76744186	co
0.87323946	0.80232558	da
0.6781609	0.72093023	dag
0.622449	0.75581395	das 
0.97333336	0.81395349	dat
0.8117647	0.87209302	dim
0.73239434	0.81395349  eg	
};\label{fig:Precision_RussianvsB}

\draw [blue,dashed] (rel axis cs:0,0) -- (rel axis cs:1,1);
\end{axis}
\draw [dashed] (c111) -- (ax2.south west);
\draw [dashed] (c222) -- (ax2.north west);
\end{tikzpicture}

\caption{Per-user Recall~\ref{fig:Recall_ImdbvsB} and Precision~\ref{fig:Precision_ImdbvsB} (RP) of PrivEdge vs fully centralised non-private State-Of-the-Art (SOA) methods for recognition of user identity (left), writer (middle) and handwritten Russian lowercase letters (right).}
\label{fig:PreRecImdb10}
\end{figure*}

\subsection{Evaluation measures}

We measure accuracy\footnote{We report accuracy using floating point training, while a trained floating point model can be quantized to integers with negligible drop in accuracy using post-training quantization~\cite{post-training}.} (i.e. precision and recall) and execution times. In the specific implementation used for validation, we defined the reconstruction dissimilarity of $\mathbf{X}$ and all of the $N$ reconstructed images as:
\begin{equation}
            \centering
                d_{i}= \sum_{l=1}^{w \times h \times c}||x(l)-\bar{x}_{i}(l)||_{2}^{2} \quad \forall i=1,2,..,N, 
\label{eq:diss_rec}     
\end{equation}
where $x(l)$ and $\bar{x}_{i}(l)$ are the $l$-th elements of image $\mathbf{X}$ and the reconstructed image $\mathbf{\bar{X}}$ by $\mbox{R}_i(\cdot)$, respectively. 

We consider precision and recall for the prediction of classes for private data of each user $u_i$ as follows:
\begin{equation}
\label{eq:rec}
    \mbox{Recall}= \frac{\mbox{TP}}{\mbox{TP}+\mbox{FN}},
\end{equation}
\begin{equation}
\label{eq:pre}
   \mbox{Precision}= \frac{\mbox{TP}}{\mbox{TP}+\mbox{FP}} ,
\end{equation}
where the prediction of class $i$ for $u_i$'s private data is a True Positive (TP); the prediction of any class different from $i$ for $u_i$'s private data is a False Negative (FN); the prediction of class $i$ for private images of other users is a  False Positive (FP); and the prediction of any class but $i$ for private images of other users is a  True Negative (TN). A conservative filter tends to be sensitive and is more likely to reject images (high FN rate), which results in low recall. A permissive filter tends to incorrectly accept images (high FP rate), which leads to a low precision.

 We also measure the prediction and local training execution times. The prediction time of PrivEdge includes an {\em offline} and {\em online} phase. The service provider and the regulator first run the offline data-independent phase to generate multiplication triplets of $\mathbf{Q}=\mathbf{U}\mathbf{V}$ for all layers (see Section~\ref{sec:prediction}). The {\em online} classification of an image evaluates $N$ reconstructors $\mbox{R}_i(\cdot)$, which can be done in parallel as their predictions are independent.    

\begin{table}
\centering
\caption{Prediction time (seconds) of the one-class reconstructor for IMDB~\cite{parkhi2015deep}, IAM~\cite{marti2002iam} and Russian lowercase letters~\cite{BaselineRussianLetter}.}
\begin{tabular}{l| r r r r}
\Xhline{3\arrayrulewidth}
\multicolumn{1}{l|}{Dataset}  & \multicolumn{1}{c }{Online Time} & \multicolumn{1}{c }{Offline Time}\\
\hline
\multicolumn{1}{l|}{IMDB} & \multicolumn{1}{c }{16} & \multicolumn{1}{c }{60}\\ 
\multicolumn{1}{l|}{IAM}  & \multicolumn{1}{c }{15} & \multicolumn{1}{c }{57}\\ 
\multicolumn{1}{l|}{Letters}  & \multicolumn{1}{c }{7} & \multicolumn{1}{c }{34}\\  
\Xhline{3\arrayrulewidth}
\end{tabular}
\label{tab:time}
\end{table}

\begin{figure}
          \centering
          \pgfplotstableread{
0.980           0.981       
0.980           0.981       
0.972           0.973        
0.968           0.970       
0.966          0.968        
0.958          0.960 
}\datatable
          \begin{tikzpicture}
          \begin{axis}[
                  name=ax1,
                  small,
                  axis lines=left, 
                  width=6cm,
                  height=5cm,
                  xtick={.75, .80, .85, .90, .95, 1},
                  xticklabels={.75, .80, .85, .90, .95, 1},
                  ytick={.90, .95, 1},
                  yticklabels={.90, .95, 1},
                  cycle list name=color list,
                  xmax=1,
                  ymax=1,
                  smooth,
                  ylabel= Precision,
                  enlarge y limits=0.01,
                  enlarge x limits=0.01,
                  xlabel={Recall},
              y tick label style={
                  /pgf/number format/.cd,
                      fixed,
                      fixed zerofill,
                      precision=2,
                  /tikz/.cd
              },
              x tick label style={
                  /pgf/number format/.cd,
                      fixed,
                      fixed zerofill,
                      precision=2,
                  /tikz/.cd}
                  ]
                  
\addplot[
    red,
    thin,
    mark=*,
    mark options={fill=red},
    visualization depends on=\thisrow{alignment} \as \alignment,
    nodes near coords, 
    point meta=explicit symbolic, 
    every node near coord/.style={anchor=\alignment} 
    ] table [
     meta index=2 
     ] {
   x             y           label       alignment

0.980           0.981        10           -40
0.980           0.981        {}           +160
0.972           0.973        {}           -40 
0.968           0.970        {}          +160
0.966          0.968        150          -40
0.958          0.960        500          -20

}; 
\label{scale-per-rec-PrivEdge}

\addplot[
    blue,
    thin,
    mark=*,
    mark options={fill=blue},
    visualization depends on=\thisrow{alignment} \as \alignment,
    nodes near coords, 
    point meta=explicit symbolic, 
    every node near coord/.style={anchor=\alignment} 
    ] table [
     meta index=2 
     ] {
   x             y           label       alignment

0.965         0.967        10             +150
0.927         0.949        {}             +160
0.842         0.939        {}            -40 
0.778         0.916        {}           +160
0.738         0.879       150           +200

}; \label{scale-per-rec-NCC}

\coordinate (c1) at (axis cs:0.980, 0.981);
\coordinate (c2) at (axis cs:0.958, 0.960);
\end{axis}

\begin{axis}[
   name=ax2,
   small,
   axis lines=left, 
   width=4cm,
   height=4cm,
   xtick={0.96,0.98},
   ytick={0.96,0.98},
   yticklabels={.96,.98},
   xticklabels={.96,.98},
   scaled ticks=false,
   xmin=0.958,
   ymin=0.960,
   at={($(ax1.south east)+(.7cm,0)$)},
   clip=true
 ]
    \addplot +[mark=*,red,mark options={fill=red}] table [
     meta index=2 
     ] {
   x             y           label       alignment

0.980           0.981        10           -40
0.980           0.981        {}           +160
0.972           0.973        {}           -40 
0.968           0.970        {}          +160
0.966          0.968        150          -40
0.958          0.960        500          -20

};

\end{axis}
\draw [dashed] (c2) -- (ax2.south west);
\draw [dashed] (c1) -- (ax2.north west);
\end{tikzpicture}

\caption{The effect of increasing the number of users on the recall and precision of PrivEdge~\ref{scale-per-rec-PrivEdge} and the $N$-class classifier~\ref{scale-per-rec-NCC} (average over 5 runs with random selection of celebrities as our users). The standard deviation is lower than 0.006 for all.  
}
\label{fig:scale}
\end{figure}
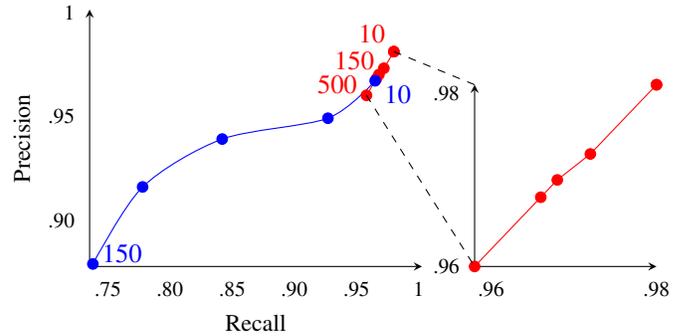

\subsection{Accuracy, timing, scalability and robustness}
The per-user recall and precision for IMDB, IAM and Russian lowercase letters are shown in Figure~\ref{fig:PreRecImdb10}. PrivEdge performs as well as the state-of-the-art methods on IMDB and Russian lowercase letters~\cite{BaselineRussianLetter}. PrivEdge can effectively distinguish handwritten texts of the users with the overall recall and precision of $92\%$ and $91\%$, respectively, although its overall precision and recall degrades by $4\%$ in comparison to the state-of-the-art non-private method on IAM~\cite{BaselineIAM}.

\begin{figure*}[t!]
\centering

\resizebox{\textwidth}{!}{ 
\begin{tabular}{N N N N N N N N N N N}
&\multicolumn{10}{c }{IMDB users}\\ 
& \multicolumn{1}{c }{\textbf{u$_1$}} & \multicolumn{1}{c }{\textbf{u$_2$}} & \multicolumn{1}{c }{\textbf{u$_3$}} & \multicolumn{1}{c }{\textbf{u$_4$}} & \multicolumn{1}{c }{\textbf{ u$_5$}} &  \multicolumn{1}{c }{\textbf{u$_6$}} & \multicolumn{1}{c }{\textbf{u$_7$}} & \multicolumn{1}{c }{\textbf{u$_8$}} & \multicolumn{1}{c }{\textbf{u$_9$}} & \multicolumn{1}{c }{\textbf{u$_{10}$}} \\
\multicolumn{1}{l}{Training data}&\multicolumn{1}{c }{\adjustimage{height=2cm,valign=m}{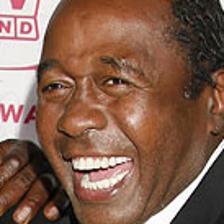}}& \multicolumn{1}{c }{\adjustimage{height=2cm,valign=m}{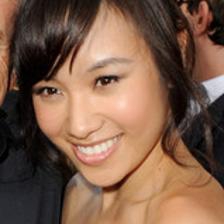}}&\multicolumn{1}{c }{\adjustimage{height=2cm,valign=m}{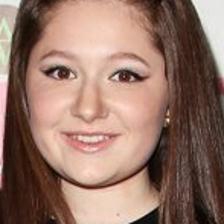}} &\multicolumn{1}{c }{\adjustimage{height=2cm,valign=m}{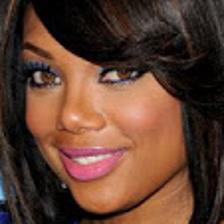}} & \multicolumn{1}{c }{\adjustimage{height=2cm,valign=m}{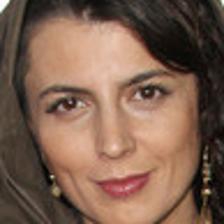}}& \multicolumn{1}{c }{\adjustimage{height=2cm,valign=m}{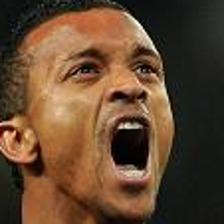}} & \multicolumn{1}{c }{\adjustimage{height=2cm,valign=m}{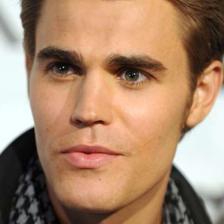}} & \multicolumn{1}{c }{\adjustimage{height=2cm,valign=m}{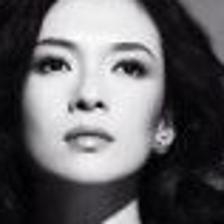}} & \multicolumn{1}{c }{\adjustimage{height=2cm,valign=m}{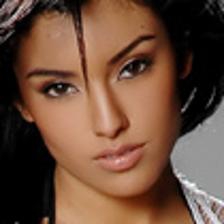}} & \multicolumn{1}{c }{\adjustimage{height=2cm,valign=m}{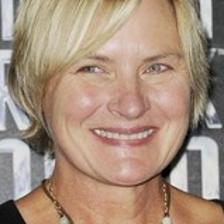}}\\
\addlinespace[0.5cm]
\multicolumn{1}{l}{Test data}&\multicolumn{1}{c }{\adjustimage{height=2cm,valign=m}{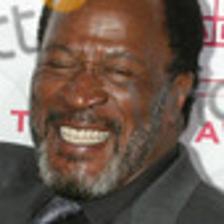}}& \multicolumn{1}{c }{\adjustimage{height=2cm,valign=m}{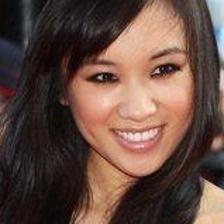}}&\multicolumn{1}{c }{\adjustimage{height=2cm,valign=m}{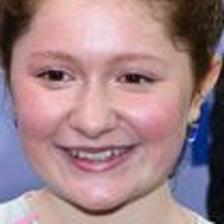}} &\multicolumn{1}{c }{\adjustimage{height=2cm,valign=m}{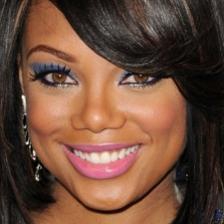}} & \multicolumn{1}{c }{\adjustimage{height=2cm,valign=m}{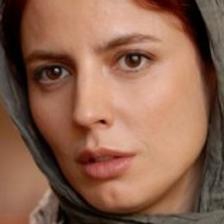}}& \multicolumn{1}{c }{\adjustimage{height=2cm,valign=m}{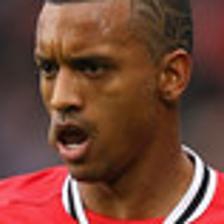}} & \multicolumn{1}{c }{\adjustimage{height=2cm,valign=m}{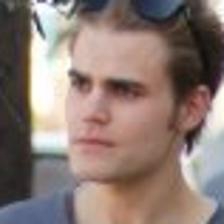}} & \multicolumn{1}{c }{\adjustimage{height=2cm,valign=m}{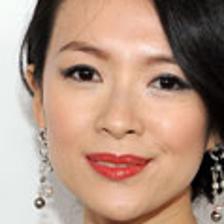}} & \multicolumn{1}{c }{\adjustimage{height=2cm,valign=m}{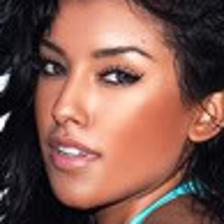}} & \multicolumn{1}{c }{\adjustimage{height=2cm,valign=m}{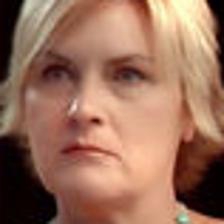}}\\

\end{tabular}
}
\caption{Example face images in training and test data of the IMDB dataset. Each user only has access to the data of one class.}
\label{tab:imdb}
\end{figure*}

The training of each one-class RAN takes 10 minutes. Table~\ref{tab:time} presents the data-independent offline and online prediction times of each one-class reconstructor for all datasets.

Prediction time measures the time of computing all linear (e.g. addition and multiplication) and non-linear (e.g. L-ReLU) functions during the prediction of a label for a given input. The number of functions and their sizes depend on the input dimension and on the size of the reconstructor, which depends on the number of layers and dimension of each layer. The dimension of the data in IMDB is $128 \times 128 \times 3$, in IAM is $128 \times 128 \times 1$ and in Russian lowercase letters is $32 \times 32 \times 3$. The reconstructors of IMDB and IAM have 14 layers, while the reconstructor of Russian lowercase letters has 12 layers.  The lower the input dimension, the fewer the layers and the lower the dimensionality of the layers of the reconstructors. Among the three datasets, Russian lowercase letters has the smallest dimension of data and, accordingly, the smallest reconstructor, resulting in the fastest prediction time (almost half of that of IMDB and IAM).
The prediction time of IMDB is 1 second more than that for IAM, as the input dimension of IMDB is 3 times larger than that of the IAM input. In general, performing multiplications is more expensive than computing L-ReLU activation functions due to the large size of matrices.

The architecture of a centralised $N$-class classifier for IMDB is chosen to have similar accuracy to PrivEdge when the number of classes are 10. This selection helps us to fairly compare the scalability of PrivEdge with a centralised $N$-class classifier by increasing the number of users. Figure~\ref{fig:scale} compares the classification part of PrivEdge (without the final privacy decision) and of a centralised $N$-class classifier for up to 500 users. Increasing the number of users has a negligible effect on the performance of the classifier, as the recall and precision drop by less than 2.2 when the number of PrivEdge users increases from 10 to 500. A possible reason for this may be advantages of $N$ one-class classifiers with regards to one $N$-class classifier when $N$ is large~\cite{mygdalis2015large,krawczyk2015usefulness}.

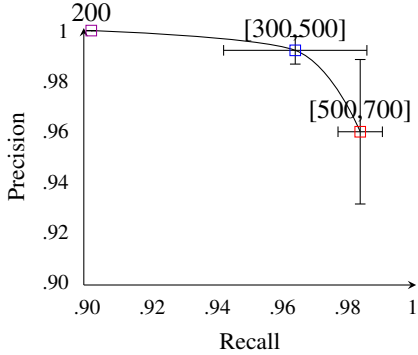
\begin{figure}
          \centering
          \begin{tikzpicture}
          \begin{axis}[
                  small,
                  axis lines=left, 
                  width=6cm,
                  height=5cm,
                  xtick={.90, .92, .94, .96, .98, 1},
                  xticklabels={.90, .92, .94, .96, .98, 1},
                  ytick={.90, .92, .94, .96, .98, 1},
                  yticklabels={.90, .92, .94, .96, .98, 1},
                  cycle list name=color list,
                  xmax=1,
                  xmin=0.9,
                  ymin=0.9,
                  smooth,
                  ylabel= Precision,
                  enlarge y limits=0.01,
                  enlarge x limits=0.01,
                  xlabel={Recall}
                  ]

\addplot[
    scatter/classes={ b={blue}, r={red}, g={green}, d={black}, c={chocolate(traditional)}, e={bostonuniversityred}, a={americanrose}, o={orange}, dg={darkgreen}, vw={violet(web)}, v={viola}, cy={cyan} , db={darkblue}, gi={giallo}, ro={rosso}  },
    scatter,
    mark=square,
    scatter src=explicit symbolic,
    nodes near coords*={\Label},
    visualization depends on={value \thisrow{label} \as \Label},
    ]
    plot [error bars/.cd, x dir = both, x explicit, y dir = both, y explicit]
    table [meta=Class, x=x,y=y,x error=ex,y error=ey] {
   x             y               ex            ey             Class    label
 0.90129035     1.               0              0               v       200
0.9638917	0.992156887      0.02198656    0.00545924           b       [300,500]
0.983826778	0.959839148     0.00681067     0.02867952           r       [500,700]
};

\end{axis}
\end{tikzpicture}
\caption{The effect of the amount of training data on the per-user recall and precision of the 10 randomly chosen users in Figure~\ref{tab:imdb}.}
\label{fig:rec-pre_IMDB}
\end{figure}

To analyse the per-user performance, we consider a set of 10 random users from IMDB (see Figure~\ref{tab:imdb}) and have repeated this experiment 5 times (Figure~\ref{fig:scale}) to verify the low standard deviation of precision and recall.

Figure~\ref{fig:rec-pre_IMDB} shows the per-user performance of the classification part of PrivEdge on 10 random users from IMDB and demonstrates that PrivEdge is robust to a training set that is not consistent in the number of images of different users. The size of the training set varies from 200 to 700.

The users whose numbers of training images are in the range $[500,700]$ achieve a higher recall than those with 200 training images. More training data helps each user $u_i$ to more accurately train their one-class classifier (i.e. smaller reconstruction dissimilarity), which enables assigning class $i$ for the images belonging to $u_i$ more often (i.e. larger TP and smaller FN). Increasing TP and decreasing FN give a higher recall (see Equation~\ref{eq:rec}). Moreover, the more accurate one-class classifier of $u_i$ can reconstruct images of those users whose classifiers are trained with few images, so it increases the possibility of FP of predicting class $i$ for images of others, thus decreases the precision (based on Equation~\ref{eq:pre}).

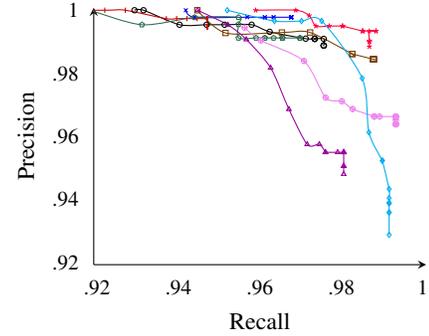
\begin{figure}[t!]
          \centering
          \begin{tikzpicture}
          \begin{axis}[
                  small,
                  axis lines=left, 
                  width=6cm,
                  height=5cm,
                  cycle list name=color list,
                  xmax=1.,
                  xmin=0.92,
                  ymin=0.92,
                  xtick={.92, .94, .96, .98, 1},
                  xticklabels={.92, .94, .96, .98, 1},
                  ytick={.92, .94, .96, .98, 1},
                  yticklabels={.92, .94, .96, .98, 1},
                  smooth,
                  ylabel= Precision,
                  enlarge y limits=0.01,
                  enlarge x limits=0.01,
                  xlabel={Recall}
                  ]
          \addplot+[black,mark=o,mark size=1]
              table[x=R0,y=P0] {per_rec_thr.txt};\label{u0}
          \addplot+[blue,mark=x,mark size=1]
              table[x=R1,y=P1] {per_rec_thr.txt};\label{u1}
          \addplot+[bostonuniversityred,mark=|,mark size=1]
              table[x=R2,y=P2] {per_rec_thr.txt};\label{u2}
          \addplot+[chocolate(traditional),mark=square,mark size=1]
              table[x=R3,y=P3] {per_rec_thr.txt};\label{u3}
          \addplot+[americanrose,mark=star,mark size=1]
              table[x=R4,y=P4] {per_rec_thr.txt};\label{u4}
          \addplot+[orange,mark=10-pointed star,mark size=1]
              table[x=R5,y=P5] {per_rec_thr.txt};\label{u5}
          \addplot+[darkgreen,mark=pentagon,mark size=1]
              table[x=R6,y=P6] {per_rec_thr.txt};\label{u6}
          \addplot+[violet(web),mark=oplus,mark size=1] 
              table[x=R7,y=P7] {per_rec_thr.txt};\label{u7}
          \addplot+[viola,mark=triangle,mark size=1]
              table[x=R8,y=P8] {per_rec_thr.txt};\label{u8}
          \addplot+[cyan,mark=diamond,mark size=1]
              table[x=R9,y=P9] {per_rec_thr.txt};\label{u9}
          \end{axis}
          \end{tikzpicture}
           \caption{The effect of the privacy threshold on the per-class recall/precision of the global filter for users $u_1$\ref{u0}, $u_2$\ref{u1}, $u_3$\ref{u2}, $u_4$\ref{u3}, $u_5$\ref{u4}, $u_6$\ref{u5}, $u_7$\ref{u6}, $u_8$\ref{u7}, $u_9$\ref{u8}, $u_{10}$\ref{u9} (IMDB dataset). The privacy threshold increases (from the top left) from 0.04 (top left) to 0.20 (bottom right).}
            \label{fig:rec-per_guard-imdb}
\end{figure}

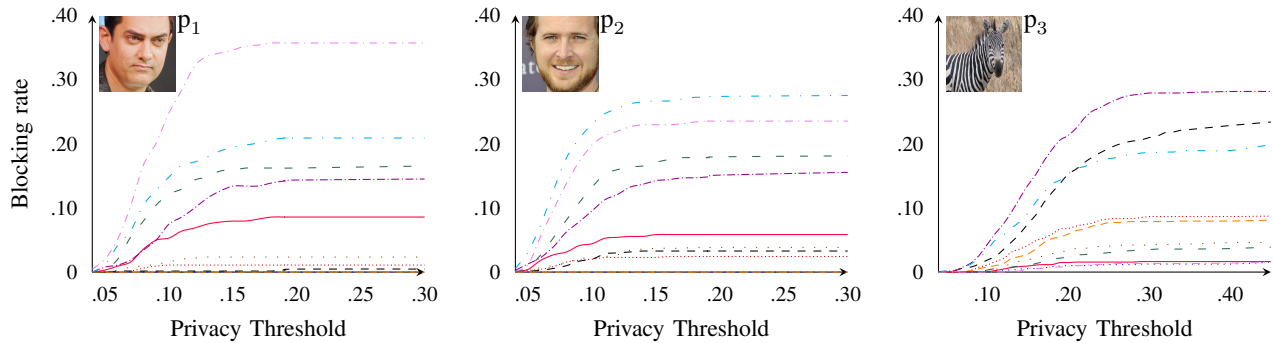
\begin{figure*}[t!]
\centering
\subfloat{
\begin{tikzpicture}
\node[inner sep=0pt] (whitehead) at (0.6,2.9)
    {\includegraphics[width=1cm]{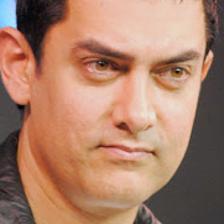}};  
\node[inner sep=0pt] (whitehead) at (1.3,3.3)
    {$\mbox{p}_1$};        
\begin{axis}[
        small,
        axis lines=left, 
        width=6cm,
        height=5cm,
        cycle list name=color list,
        ymin=0,
        ymax=0.4,
        smooth,
        xtick={.05, .10, .15, .20, .25, .30},
        xticklabels={.05, .10, .15, .20, .25, .30},
        ytick={0, .10, .20, .30, .40},
        yticklabels={0, .10, .20, .30, .40},
        ylabel = {Blocking rate},
        enlarge y limits=0.0,
        enlarge x limits=0.0,
        xlabel={Privacy Threshold}
        ]
\addplot+[black,dashed]
    table[x=u,y=p0] {amir_blocked.txt};\label{u0_b}
\addplot+[blue,dotted]
    table[x=u,y=p1] {amir_blocked.txt};\label{u1_b}
\addplot+[bostonuniversityred,densely dotted]
    table[x=u,y=p2] {amir_blocked.txt};\label{u2_b}
\addplot+[chocolate(traditional),loosely dotted]
    table[x=u,y=p3] {amir_blocked.txt};\label{u3_b}
\addplot+[americanrose, solid]
    table[x=u,y=p4] {amir_blocked.txt};\label{u4_b}
\addplot+[orange, densely dashed]
    table[x=u,y=p5] {amir_blocked.txt};\label{u5_b}
\addplot+[darkgreen, loosely dashed]
    table[x=u,y=p6] {amir_blocked.txt};\label{u6_b}
\addplot+[violet(web), dashdotted]
    table[x=u,y=p7] {amir_blocked.txt};\label{u7_b}
\addplot+[viola, densely dashdotted]
    table[x=u,y=p8] {amir_blocked.txt};\label{u8_b}
\addplot+[cyan,loosely dashdotted]
    table[x=u,y=p9] {amir_blocked.txt};\label{u9_b}

\end{axis}
\end{tikzpicture}
}
\subfloat{
\begin{tikzpicture}
\node[inner sep=0pt] (whitehead) at (0.6,2.9)
    {\includegraphics[width=1cm]{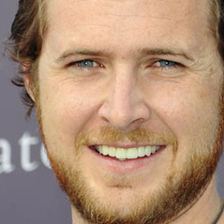}};  
\node[inner sep=0pt] (whitehead) at (1.3,3.3)
    {$\mbox{p}_2$};      
\begin{axis}[
        small,
        axis lines=left, 
        width=6cm,
        height=5cm,
        cycle list name=color list,
        ymin=0,
        ymax=0.4,
        smooth,
        enlarge y limits=0.0,
        enlarge x limits=0.0,
        xtick={.05, .10, .15, .20, .25, .30},
        xticklabels={.05, .10, .15, .20, .25, .30},
        ytick={0, .10, .20, .30, .40},
        yticklabels={0, .10, .20, .30, .40},
        xlabel={Privacy Threshold}
        ]
\addplot+[black,dashed]
    table[x=u,y=p0] {buckley_blocked.txt};
\addplot+[blue,dotted]
    table[x=u,y=p1] {buckley_blocked.txt};
\addplot+[bostonuniversityred,densely dotted]
    table[x=u,y=p2] {buckley_blocked.txt};
\addplot+[chocolate(traditional),loosely dotted]
    table[x=u,y=p3] {buckley_blocked.txt};
\addplot+[americanrose, solid]
    table[x=u,y=p4] {buckley_blocked.txt};
\addplot+[orange, densely dashed]
    table[x=u,y=p5] {buckley_blocked.txt};
\addplot+[darkgreen, loosely dashed]
    table[x=u,y=p6] {buckley_blocked.txt};
\addplot+[violet(web), dashdotted]
    table[x=u,y=p7] {buckley_blocked.txt};
\addplot+[viola, densely dashdotted]
    table[x=u,y=p8] {buckley_blocked.txt};
\addplot+[cyan,loosely dashdotted]
    table[x=u,y=p9] {buckley_blocked.txt};
\end{axis}
\end{tikzpicture}

}
\subfloat{
\begin{tikzpicture}
\node[inner sep=0pt] (whitehead) at (0.6,2.9)
    {\includegraphics[width=1cm]{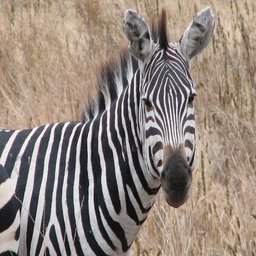}}; 
\node[inner sep=0pt] (whitehead) at (1.3,3.3)
    {$\mbox{p}_3$};      
\begin{axis}[
        small,
        axis lines=left, 
        width=6cm,
        height=5cm,
        cycle list name=color list,
        ymin=0,
        ymax=0.4,
        smooth,
        enlarge y limits=0.0,
        enlarge x limits=0.0,
        xlabel={Privacy Threshold},
        xtick={.10, .20, .30, .40},
        xticklabels={.10, .20, .30, .40},
        ytick={0, .10, .20, .30, .40},
        yticklabels={0, .10, .20, .30, .40},
        ]
\addplot+[black,dashed]
    table[x=u,y=p0] {zebra_blocked.txt};
\addplot+[blue,dotted]
    table[x=u,y=p1] {zebra_blocked.txt};
\addplot+[bostonuniversityred,densely dotted]
    table[x=u,y=p2] {zebra_blocked.txt};
\addplot+[chocolate(traditional),loosely dotted]
    table[x=u,y=p3] {zebra_blocked.txt};
\addplot+[americanrose, solid]
    table[x=u,y=p4] {zebra_blocked.txt};
\addplot+[orange, densely dashed]
    table[x=u,y=p5] {zebra_blocked.txt};
\addplot+[darkgreen, loosely dashed]
    table[x=u,y=p6] {zebra_blocked.txt};
\addplot+[violet(web), dashdotted]
    table[x=u,y=p7] {zebra_blocked.txt};
\addplot+[viola, densely dashdotted]
    table[x=u,y=p8] {zebra_blocked.txt};
\addplot+[cyan,loosely dashdotted]
    table[x=u,y=p9] {zebra_blocked.txt};
\end{axis}
\end{tikzpicture}
}
 \caption{The cumulative blocking rate of non-private images (two non-registered faces $\mbox{p}_1$ and $\mbox{p}_2$, and Zebra $\mbox{p}_3$) by one-class reconstructors, $\mbox{R}_1$\ref{u0_b}, $\mbox{R}_2$\ref{u1_b}, $\mbox{R}_3$\ref{u2_b}, $\mbox{R}_4$\ref{u3_b}, $\mbox{R}_5$\ref{u4_b}, $\mbox{R}_6$\ref{u5_b}, $\mbox{R}_7$\ref{u6_b}, $\mbox{R}_8$\ref{u7_b}, $\mbox{R}_9$\ref{u8_b}, $\mbox{R}_{10}$\ref{u9_b}. These images are uploaded by a user who did not participate in the training. 
 }
  \label{fig:guard-faces}
\end{figure*}

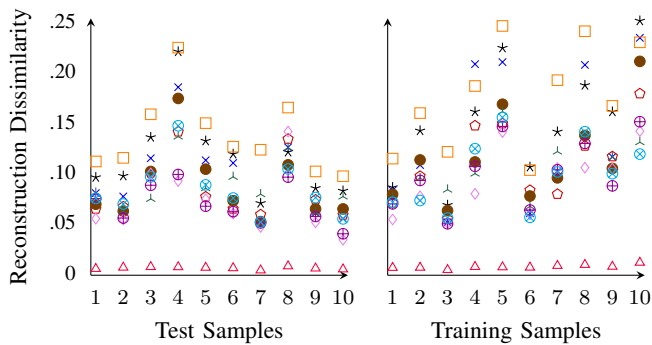
\begin{figure}[t!]
          \centering

          \subfloat{
          \begin{tikzpicture}
          \begin{axis}[cycle list name=color list,
                  small,
                  axis lines=left, 
                  xtick={1,2,...,10},
                  ytick={0,.05,.10,.15,.20,.25},
                  yticklabels={0,.05,.10,.15,.20,.25},
                  width=5cm,
                  height=5cm,
                  ymin=0.00,
                  ymax=0.25,
                  enlarge y limits=0.01,
                  enlarge x limits=0.02,
                  xlabel={Test Samples},
                  ylabel={Reconstruction Dissimilarity},
                  xtick={1,2,3,4,5,6,7,8,9,10}
                  ]
          \addplot+[black,only marks,mark=star]
              table[x=q,y=p0] {usr4test.txt};\label{u0_rec}
          \addplot+[blue,only marks,mark=x]
              table[x=q,y=p1] {usr4test.txt};\label{u1_rec}
          \addplot+[bostonuniversityred,only marks,mark=pentagon]
              table[x=q,y=p2] {usr4test.txt};\label{u2_rec}
          \addplot+[chocolate(traditional),only marks,mark=*]
              table[x=q,y=p3] {usr4test.txt};\label{u3_rec}
          \addplot+[americanrose,only marks,mark=triangle]
              table[x=q,y=p4] {usr4test.txt};\label{u4_rec}
          \addplot+[orange,only marks, mark=square]
              table[x=q,y=p5] {usr4test.txt};\label{u5_rec}
          \addplot+[darkgreen,only marks,mark=Mercedes star]
              table[x=q,y=p6] {usr4test.txt};\label{u6_rec}
          \addplot+[violet(web),only marks,mark=diamond]
              table[x=q,y=p7] {usr4test.txt};\label{u7_rec}
           \addplot+[viola,only marks,mark=oplus]
               table[x=q,y=p8] {usr4test.txt};\label{u8_rec}
           \addplot+[cyan,only marks,mark=otimes]
               table[x=q,y=p9] {usr4test.txt};\label{u9_rec}
           \end{axis}
          \end{tikzpicture}
          } \hspace*{-0.9em}        
           \subfloat{
          \begin{tikzpicture}
          \begin{axis}[cycle list name=color list,
                  small,
                  axis lines=left, 
                  xtick={1,2,...,10},
                  width=5cm,
                  height=5cm,
                  ymin=0.00,
                  enlarge y limits=0.01,
                  enlarge x limits=0.02,
                  xlabel={Training Samples},
                  yticklabels={},
                  ]
          \addplot+[black,only marks,mark=star]
              table[x=q,y=p0] {usr4train.txt};\label{u0_rec}
          \addplot+[blue,only marks,mark=x]
              table[x=q,y=p1] {usr4train.txt};\label{u1_rec}
          \addplot+[bostonuniversityred,only marks,mark=pentagon]
              table[x=q,y=p2] {usr4train.txt};\label{u2_rec}
          \addplot+[chocolate(traditional),only marks,mark=*]
              table[x=q,y=p3] {usr4train.txt};\label{u3_rec}
          \addplot+[americanrose,only marks,mark=triangle]
              table[x=q,y=p4] {usr4train.txt};\label{u4_rec}
          \addplot+[orange,only marks,mark=square]
              table[x=q,y=p5] {usr4train.txt};\label{u5_rec}
          \addplot+[darkgreen,only marks,mark=Mercedes star]
              table[x=q,y=p6] {usr4train.txt};\label{u6_rec}
          \addplot+[violet(web),only marks,mark=diamond]
              table[x=q,y=p7] {usr4train.txt};\label{u7_rec}
           \addplot+[viola,only marks,mark=oplus]
               table[x=q,y=p8] {usr4train.txt};\label{u8_rec}
           \addplot+[cyan,only marks,mark=otimes]
               table[x=q,y=p9] {usr4train.txt};\label{u9_rec}
           \end{axis}
          \end{tikzpicture}
          }
           \caption{The reconstruction dissimilarities of 10 test and training faces of user $u_5$ by 10 one-class reconstructors, $\mbox{R}_1$\ref{u0_rec}, $\mbox{R}_2$\ref{u1_rec}, $\mbox{R}_3$\ref{u2_rec}, $\mbox{R}_4$\ref{u3_rec}, $\mbox{R}_5$\ref{u4_rec}, $\mbox{R}_6$\ref{u5_rec}, $\mbox{R}_7$\ref{u6_rec}, $\mbox{R}_8$\ref{u7_rec}, $\mbox{R}_9$\ref{u8_rec}, $\mbox{R}_{10}$\ref{u9_rec}.}
            \label{fig:rec-err}
\end{figure}
\subsection{Privacy protection}

The lower the privacy threshold the more images will be classified as non-private, thus increasing the number of FNs. Figure~\ref{fig:rec-per_guard-imdb} shows the per-user precision and recall of blocking private images by each $\mbox{R}_i(\cdot)$ when changing the privacy threshold from $0.04$ to $0.20$ with intervals of length $0.01$. It can be seen that increasing the privacy threshold increases, as expected, the recall.

We also consider the percentage of blocked non-private images due to each individual one-class reconstructor, as a function of the privacy threshold. The percentage of blocking non-private images by the PrivEdge classifier due to each one-class $\mbox{R}_i(\cdot)$ is the ratio between the number of non-private images that are classified as private of user $u_i$ and the total number of non-private images. We consider 3 sets of images, $\mbox{p}_1, \mbox{p}_2$ and $\mbox{p}_3$, from $2$ non-private classes (i.e. different or similar to the  privacy-sensitive images). Sets $\mbox{p}_1$ and $\mbox{p}_2$ comprise the faces of two randomly chosen celebrities from the IMDB dataset who have not participated in the training of the filter. The other set, $\mbox{p}_3$, contains images from the ImageNet~\cite{russakovsky2015imagenet} Zebra class.

In Figure~\ref{fig:guard-faces}, we show the values of the privacy thresholds that cause the blocking of $0\%$ to $100\%$ of non-private images. For a specific privacy threshold, the number of blocked non-private images of faces is larger than that of blocked non-private Zebra images, as non-private images from $\mbox{p}_1$ and $\mbox{p}_2$ are more similar to the private images in the training data than Zebra images. By adding the percentage of blocked non-private images by the filter due to all of the one-class reconstructors, we see all non-private images of faces are blocked when the privacy threshold is $0.3$. However, the same occurs for Zebra images with a higher privacy threshold of $0.45$. This confirms that trained one-class classifiers reconstruct images that are similar to the training images with lower reconstruction dissimilarities than images with different structures.

Figure~\ref{fig:scale} and Figure~\ref{fig:guard-faces} show that PrivEdge is accurate in classifying and detecting private images uploaded by others (i.e. it is conservative), yet it is permissive of a user uploading their own private images and non-private images. Therefore, the filter does not interfere with image-sharing when there are no concerns of violating the privacy of others.

Figure~\ref{fig:rec-err} compares the reconstruction dissimilarities of 10 one-class reconstructors for 10 random test and training samples of a specific user's face, $u_5$. As expected, $\mbox{R}_5$ reconstructs  both test and training sample faces of user $u_5$ better than other reconstructors. 

\subsection{Visual assessment and robustness to compression}
\begin{figure*}[t!]
  \includegraphics[width=\linewidth]{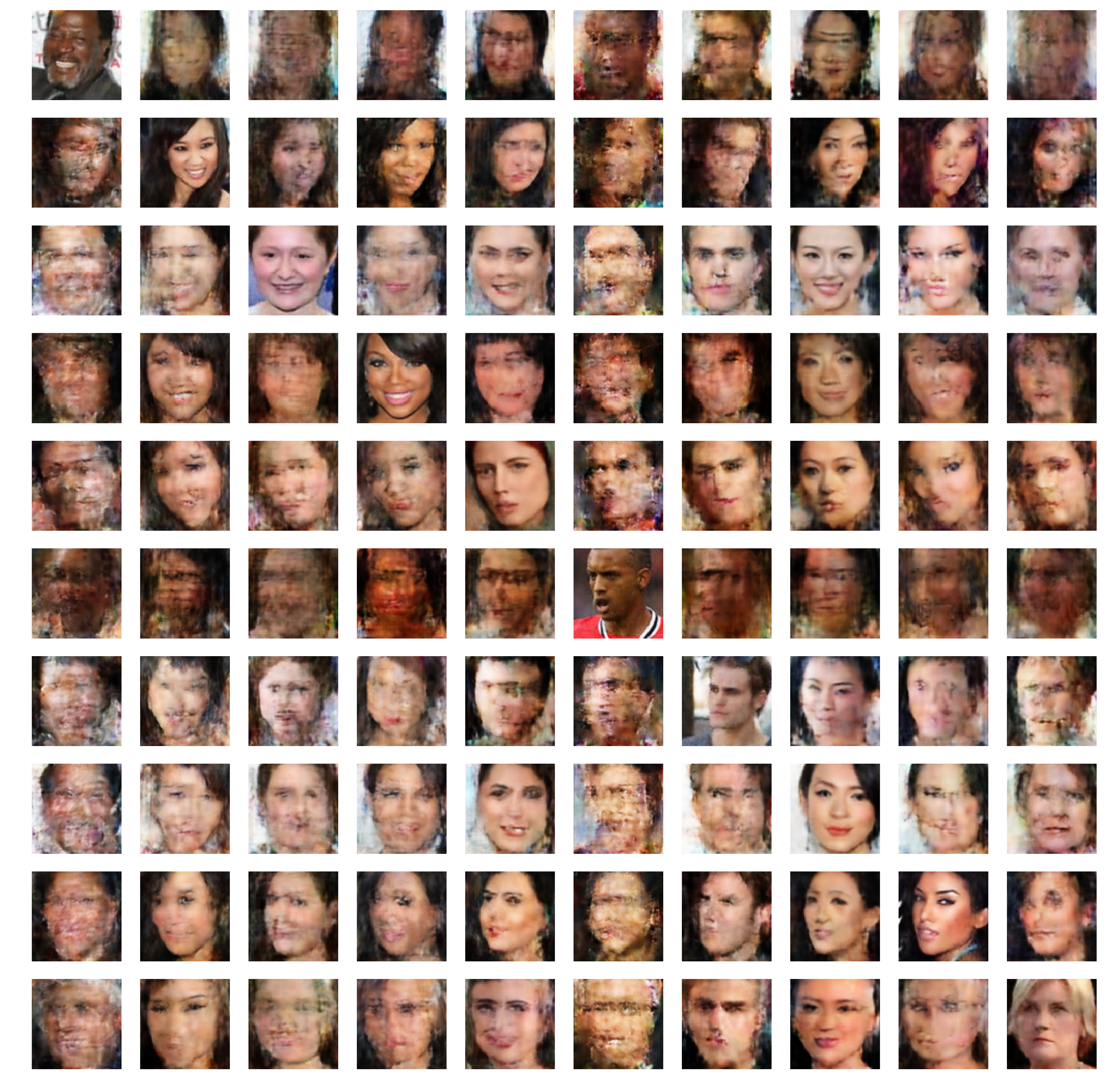}
  \caption{Examples of reconstructed private images. Each row shows the reconstruction of the (test) face of each user in Figure~\ref{tab:imdb}, by all the 10 one-class reconstructors. The face of each user is reconstructed by their corresponding one-class reconstructor (diagonal) better than other one-class reconstructors.}
  \label{fig:rec_imdb_org}
\end{figure*}

To better illustrate both the reconstruction dissimilarity of each trained reconstructor for images of different classes and the advantages of one-class decomposition from an accuracy perspective, we visualise the reconstructed images. Figure~\ref{fig:rec_imdb_org} shows the reconstructed images by 10 one-class reconstructors for images of the 10 users in Figure~\ref{tab:imdb}. As expected, the best reconstruction of each image is produced by its corresponding reconstructor. 
\begin{figure*}[t!]
  \includegraphics[width=\linewidth]{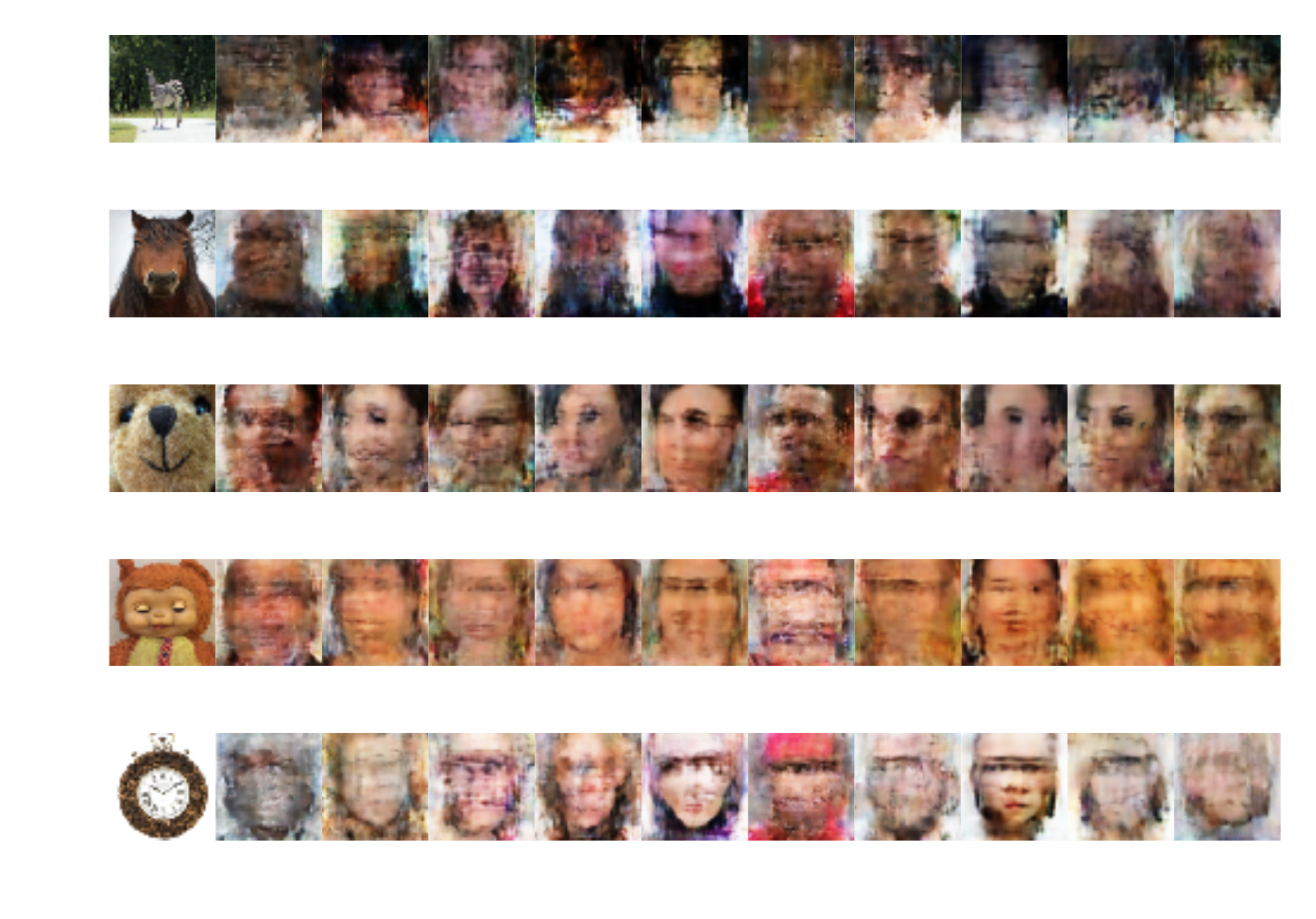}
  \caption{Examples of reconstructed non-private images. The first column includes original images of Zebra, Horse (face), Teddy Bear (face), plastic Teddy Bear (face) and Wall Clock. The second to last columns show the reconstructed images by 10 one-class reconstructors trained on faces.}
  \label{fig:rec_zebra}
\end{figure*}
In addition, we visualise the reconstruction of several non-private images by 10 one-class reconstructors of the filter in Figure~\ref{fig:rec_zebra}. This shows the ability of PrivEdge to distinguish private images from non-private images based on the reconstruction dissimilarities.

Finally, as service providers use different image transcoding techniques and users themselves may compress images at different qualities, we test the robustness of the prediction to different encodings and compression ratios. 
To this end, we evaluate the robustness of PrivEdge, trained with original quality images, to images compressed at different JPEG quality levels.
Figure~\ref{fig:jpeg_rec-pre_IMDB_avg_std} shows that the impact of JPEG compression on the performance of PrivEdge is negligible as the accuracy only slightly changes when degrading the encoding quality from 100 to 2.

\newcommand{\markerfour}{\raisebox{0.5pt}{\tikz{\node[blue, draw,scale=0.4,regular polygon, regular polygon sides=4,fill=none](){};}}}

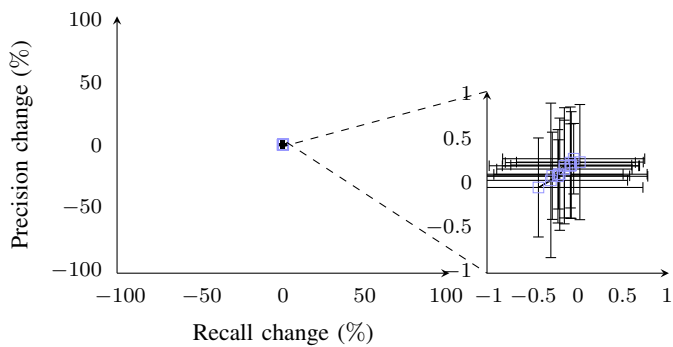
\begin{figure}
          \centering
          \begin{tikzpicture}
          \begin{axis}[
                  name=ax1,
                  small,
                  axis lines=left, 
                  width=6cm,
                  height=5cm,
                  cycle list name=color list,
                  xmax=100,
                  xmin=-100,
                  ymin=-100,
                  ymax=100,
                  smooth,
                  ylabel= Precision change ($\%$),
                  enlarge y limits=0.01,
                  enlarge x limits=0.01,
                  xlabel={Recall change ($\%$)},
                  ]

\addplot[
    scatter/classes={ b2={blue!40}, b5={blue!40}, b10={blue!40}, b15={blue!40}, b20={blue!40}, b30={blue!40}, b40={blue!40}, b50={blue!40}, b60={blue!40}, b70={blue!40}, b80={blue!40}, b90={blue!40}, b100={blue!40}},
    scatter,
    mark=square,
    scatter src=explicit symbolic,
    ]
    plot [error bars/.cd, x dir = both, x explicit, y dir = both, y explicit]
    table [meta=Class, x=x,y=y,x error=ex,y error=ey] {
   x             y               ex            ey             Class    
 -0.304139739	0.021821966	    0.860422198	0.866900365         b2
-0.204341202	0.090931156	    0.980094674	0.629707371         b5
-0.443277598	-0.057851755	1.171740897	0.555606719         b10
-0.2246216	    0.06762318	    1.00738202	0.523921777         b15
-0.152545895	0.185671541	    0.842981234	0.651622267         b20
-0.288357812	0.073275227	    0.867996687	0.489554379         b30
-0.2186999	    0.088969537	    0.724670341	0.388542229         b40
-0.085355961	0.221550714	    0.729168124	0.623860695         b50
-0.048195504	0.264795812	    0.796810074	0.39552435          b60
0.021320679	    0.226870236	    0.710862256	0.646286238         b70
-0.083486338	0.184319413	    0.764770186	0.471994349         b80
-0.075239639	0.193898321	    0.677989501	0.598756078         b90
-0.15225113	    0.14879728	    0.758819168	0.548560461         b100
};
\coordinate (c11111) at (axis cs:-2, -2);
\coordinate (c22222) at (axis cs:2, 2);
\end{axis}

         \begin{axis}[
                  name=ax2,
                  small,
                  axis lines=left, 
                  width=4cm,
                  height=4cm,
                  cycle list name=color list,
                  xmax=1,
                  xmin=-1,
                  ymin=-1,
                  ymax=1,
                  smooth,
                  enlarge y limits=0.01,
                  enlarge x limits=0.01,
                  at={($(ax1.south east)+(0.5cm,0)$)},
                  ]

\addplot[
    scatter/classes={ b2={blue!40}, b5={blue!40}, b10={blue!40}, b15={blue!40}, b20={blue!40}, b30={blue!40}, b40={blue!40}, b50={blue!40}, b60={blue!40}, b70={blue!40}, b80={blue!40}, b90={blue!40}, b100={blue!40}},
    scatter,
    mark=square,
    scatter src=explicit symbolic,
    ]
    plot [error bars/.cd, x dir = both, x explicit, y dir = both, y explicit]
    table [meta=Class, x=x,y=y,x error=ex,y error=ey] {
   x             y               ex            ey             Class    
 -0.304139739	0.021821966	    0.860422198	0.866900365         b2
-0.204341202	0.090931156	    0.980094674	0.629707371         b5
-0.443277598	-0.057851755	1.171740897	0.555606719         b10
-0.2246216	    0.06762318	    1.00738202	0.523921777         b15
-0.152545895	0.185671541	    0.842981234	0.651622267         b20
-0.288357812	0.073275227	    0.867996687	0.489554379         b30
-0.2186999	    0.088969537	    0.724670341	0.388542229         b40
-0.085355961	0.221550714	    0.729168124	0.623860695         b50
-0.048195504	0.264795812	    0.796810074	0.39552435          b60
0.021320679	    0.226870236	    0.710862256	0.646286238         b70
-0.083486338	0.184319413	    0.764770186	0.471994349         b80
-0.075239639	0.193898321	    0.677989501	0.598756078         b90
-0.15225113	    0.14879728	    0.758819168	0.548560461         b100
};
\end{axis}
\draw [dashed] (c22222) -- (ax2.south west);
\draw [dashed] (c11111) -- (ax2.north west);
\end{tikzpicture}
\caption{The effect of JPEG compression on the per-class recall and precision of the filter for quality level 2, 5, 10, 15, 20, 30, 40, 50, 60, 70, 80, 90 and 100 (IMDB dataset). It represents the average~\protect\markerfour ~and standard deviation~{\protect\raisebox{2pt}{\protect\tikz \protect\draw[black,line width=1] (0,0) -- (0.3,0);}} of differences of precision and recall of corresponding JPEG quality from  precision and recall of the original quality.}
\label{fig:jpeg_rec-pre_IMDB_avg_std}
\end{figure}

\section{Conclusion}
\label{sec:conclude}

We proposed \emph{PrivEdge}, a technique for privacy-preserving MLaaS where training data, model parameters, and prediction data of each user remain private during both training and prediction. In PrivEdge, users locally and privately train an instance of a one-class Reconstructive Adversarial Network (RAN) as a one-class classifier. This one-class RAN describes a set of private pre-selected images and learns to reconstruct its training data. Then, for the prediction phase, a multi-class classifier is aggregated in the cloud by leveraging a third party (a regulator) that aids in the computation while learning nothing about user data.

Future work includes extending the validation of PrivEdge to different types of image manipulations and compositions, such as private adversarial examples~\cite{Li2019}.

\ifCLASSOPTIONcaptionsoff
  \newpage
\fi

\end{document}